\documentclass[printer]{aa}  
\pdfoutput=1
\usepackage{graphicx}
\usepackage{natbib}
\usepackage{float}
\usepackage{color}
\usepackage{multirow}
\usepackage[para,online,flushleft]{threeparttable}
\usepackage{outlines}
\usepackage{txfonts}
\usepackage[]{hyperref}
\hypersetup{
  colorlinks,
  citecolor=blue,
  linkcolor=red,
  urlcolor=blue}
\usepackage[dvipsnames]{xcolor}

\bibpunct{(}{)}{;}{a}{}{,} 
\newcommand{\psr}{PSR~J1910$-$5959A}
\def\lapp{\ifmmode\stackrel{<}{_{\sim}}\else$\stackrel{<}{_{\sim}}$\fi}
\def\gapp{\ifmmode\stackrel{>}{_{\sim}}\else$\stackrel{<}{_{\sim}}$\fi}

\begin{document} 

   \title{\psr: A rare gravitational laboratory for testing white dwarf models}
   \titlerunning{\psr: A rare gravitational laboratory for testing white dwarf models}
\author{ A.~Corongiu \inst{\ref{inaf}} \and  
V.~Venkatraman~Krishnan\inst{\ref{mpifr}\dagger }\and 
P.~C.~C.~Freire\inst{\ref{mpifr}}\and 
M.~Kramer\inst{\ref{mpifr},\ref{manchester}}\and 
A.~Possenti\inst{\ref{inaf}}\and 
M.~Geyer\inst{\ref{sarao}, \ref{uct}} \and
A. Ridolfi\inst{\ref{inaf},\ref{mpifr}}
F.~Abbate\inst{\ref{mpifr}}\and
M.~Bailes\inst{\ref{swinburne},\ref{ozgrav}} \and
E.~D.~Barr\inst{\ref{mpifr}}\and 
V. Balakrishnan\inst{\ref{mpifr}}\and
S.~Buchner\inst{\ref{sarao}} \and
D.~J.~Champion\inst{\ref{mpifr}}\and
W. Chen\inst{\ref{mpifr}}\and
B.~V.~Hugo\inst{\ref{sarao},\ref{rhodesphysics}} \and
A.~Karastergiou \inst{\ref{oxford}} \and
A.~G.~Lyne \inst{\ref{manchester}} \and
R.~N.~Manchester\inst{\ref{csiro}} \and
P.~V.~Padmanabh\inst{\ref{mpifr}, \ref{aei}} \and
A.~Parthasarathy\inst{\ref{mpifr}} \and
S.~M.~Ransom\inst{\ref{nrao}}\and
J.~M.~Sarkissian\inst{\ref{pks}} \and
M.~Serylak\inst{\ref{skao}, \ref{uwc}} \and
W.~van~Straten\inst{\ref{aut}}  
}
  \institute{ 
    INAF-Osservatorio Astronomico di Cagliari, via della Scienza 5, 09044 Selargius, Italy \label{inaf} 
    \and   
    Max-Planck-Institut f\"{u}r Radioastronomie, Auf dem H\"{u}gel 69, D-53121 Bonn, Germany\label{mpifr} 
    \and
    Jodrell Bank Centre for Astrophysics., School of Physics and Astronomy., Univ. of Manchester, Manchester., M13 9PL, UK \label{manchester}
    \and
    South African Radio Astronomy Observatory, 2 Fir Street, Black River Park, Observatory 7925, South Africa\label{sarao} 
    \and 
    Department of Astronomy, University of Cape Town, Rondebosch, Cape Town, 7700, South Africa \label{uct}
    \and
    Centre for Astrophysics and Supercomputing, Swinburne University of Technology, PO Box 218, Hawthorn, Vic, 3122, Australia
    \label{swinburne}
    \and
    The ARC Centre of Excellence for Gravitational Wave Discovery (OzGrav)
    \label{ozgrav}
    \and
    Department of Physics and Electronics, Rhodes University, PO Box 94, Grahamstown 6140, South Africa\label{rhodesphysics}
    \and
    Astrophysics, University of Oxford, Denys Wilkinson Building, Keble Road, Oxford OX1 3RH, UK \label{oxford}
    \and
    Australia Telescope National Facility, CSIRO, Space and Astronomy, Epping NSW 1710, Australia
    \label{csiro}
    \and
    Max-Planck-Institut f\"{u}r Gravitationsphysik (Albert-Einstein-Institut), D-30167 Hannover, Germany
    \label{aei}
    \and
    National Radio Astronomy Observatory, 520 Edgemont Rd., Charlottsville, VA, 22903, USA\label{nrao}    
    \and
    Australia Telescope National Facility, CSIRO, Space and Astronomy, Parkes NSW 2870, Australia
    \label{pks}
    \and
    SKA Observatory, Jodrell Bank, Lower Withington, Macclesfield, SK11 9FT, United Kingdom
    \label{skao}
    \and
    Department of Physics and Astronomy, University of the Western Cape, Bellville, Cape Town, 7535, South Africa
    \label{uwc}
    \and
    Institute for Radio Astronomy \& Space Research,
    Auckland University of Technology, 
    Auckland 1142, New Zealand
    \label{aut}
  \\
   $\dagger$\email{vkrishnan@mpifr-bonn.mpg.de} 
}
   \date{Received --; accepted --}

\abstract
{PSR J1910$-$5959A is a binary millisecond pulsar in a $0.837$ day circular orbit around a helium white dwarf (HeWD) companion. The position of this pulsar is 6.3 arcminutes ($\sim$74 core radii) away from the optical centre of the globular cluster (GC) NGC6752. Given the large offset, the association of the pulsar with the GC has been debated.}
{We aim to obtain precise measurements of the masses of the stars in the system along with secular orbital parameters, which will help identify if the system belongs to the GC. }
{We have made use of archival Parkes 64 m `Murriyang' telescope data and carried out observations with the MeerKAT telescope with different backends and receivers over the last two decades. Pulse times of arrival were obtained from these using standard pulsar data reduction techniques and analysed using state-of-the-art Bayesian pulsar timing techniques. We also performed an analysis of the pulsar's total intensity and polarisation profile to understand the interstellar scattering along the line of sight, and we determined the pulsar's geometry by fitting the rotating vector model to the polarisation data.}
{We obtain precise measurements of several post-Keplerian parameters: 
the range, $r=0.202(6)\,T_\odot$, and shape, $s=0.999823(4),$ of the Shapiro delay, from which we infer: the orbital inclination to be  $88.9^{+0.15}_{-0.14} \deg$; the masses of the pulsar and the companion to be $1.55(7) M_{\odot}$ and $0.202(6) M_{\odot}$, respectively; a secular change in the orbital period $\dot{P}_{\rm b}=-53^{+7.4}_{-6.0} \times 10^{-15}$\,s\,s$^{-1}$ that proves the GC association; and a secular change in the projected semi-major axis of the pulsar, $\dot{x}= -40.7^{+7.3}_{-8.2}\times10^{-16}$\,s\,s$^{-1}$,  likely caused by the spin--orbit interaction from a misaligned HeWD spin,
at odds with the likely isolated
binary evolution of the system. We also discuss some theoretical models for the structure and evolution of white dwarfs in neutron star--white dwarf binaries, using \psr's companion as a test bed.
}
{PSR\,J1910-5959A is a rare system for which several parameters of both the pulsar and the HeWD companion can be accurately measured. As such, it is a test bed for discriminating between alternative models of HeWD structure and cooling.}

\maketitle

\section{Introduction}
\label{sec:intro}

Pulsars are rapidly rotating neutron stars \citep[NSs;][]{Gold1968}, whose radio signals are often observed as a periodic sequence of pulses. The periodicity of the observed pulses has a stability comparable, in some cases, with that of atomic clocks, thus making them perfect tools for investigating a wide range of fields in physics and astrophysics. In particular, observations of pulsars in close binary systems allow accurate measurements of the Keplerian orbital motion and its deviations due to relativistic effects (see e.g. \citealt{ksm+21}), from which information on the binary companion can also be obtained.

The binary pulsar \psr\, (PSRA in subscripts), located in the outskirts of the core-collapsed globular cluster (GC) NGC6752, has a spin period of $3.26$ milliseconds and is in a 0.837\,day circular orbit around a $\sim0.2M_\odot$ companion \citep{dlm+01}. The precise position derived from the timing of the pulsar \citep{dpf+02} immediately demonstrated an important peculiarity of this object: \psr's angular distance from the centre of NGC6752 is 6.37\,arcminutes (\citealt{dpf+02}, \citealt{cpl+06}), corresponding to 1.4 half-mass radii and 0.11 tidal radii (according to the values reported by \citealt{hbsb20}). The offset of the position of \psr\, with respect to the centre of NGC6752  is difficult to explain as most NSs lie much closer to the cores of their clusters owing to mass segregation. The system could have been ejected from those inner regions by a close encounter with another star system; however, such events tend to increase the orbital eccentricities of the binaries involved \citep{1992RSPTA.341...39P}. Strangely, the \psr\, system has a very low eccentricity, showing no sign of such a close encounter. A possibility is that the system was ejected by a not-so-close close encounter with a binary black hole at the centre of the cluster \citep{2002ApJ...570L..85C}. As an alternative, some authors have called into question its association with NGC6752 \citep{bkkv06}.

\psr's aforementioned companion has been identified in optical observations with the \textit{Hubble} Space Telescope (HST; see \citealt{bvkh03}) and the European Southern Observatory (ESO) Very Large Telescope (VLT; see \citealt{fpsd03}). These observations confirm that the companion is a helium white dwarf (HeWD) star and provide consistent estimates of its mass, $M_{\rm C} \geq0.185\,M_\odot$ and $M_{\rm C}=0.17-0.20\,M_\odot$, respectively, and surface temperature,  $T_{\rm eff} = 11,000-16,000$\,K and $T_{\rm eff} = 10,000-12,000$\,K, respectively.

After the identification, \cite{cfpd06} and \cite{bkkv06} made orbital phase-resolved spectroscopic optical observations of the companion with the ESO VLT.
They combined its radial velocity curve with the orbital parameters derived from the pulsar
timing to obtain the system mass ratio, $q\equiv M_{\rm P}/M_{\rm C}=7.49\pm0.64$
and $q=7.36\pm0.25$, respectively. These values, once combined with the estimates of the companion mass from the aforementioned optical observations, imply pulsar masses of
$1.40^{+0.16}_{-0.10}\,M_\odot$ and $1.36 \pm 0.08\,M_\odot$, respectively.
While these values are in agreement, the measurements
of the systemic radial velocity,
$V_{\rm R}=-28.1\pm4.9$\,km\,s$^{-1}$
and $V_{\rm R}=-18\pm6$\,km\,s$^{-1}$, respectively, are only barely consistent at the 1$\sigma$ level; the former is in better agreement
with the heliocentric radial velocity of NGC6752 of $-26.28\pm0.16$\,km\,s$^{-1}$
\citep{vas19}.

Moreover, using their optical observations, \cite{bkkv06} inferred a distance
$D=3.1\pm0.7$\,kpc to the system, which is marginally inconsistent with the best distance estimate for
NGC6752 of 4.14\,kpc \citep{gbc+03} that was available at that time.
Starting from this inconsistency, \cite{bkkv06} proposed the non-association of \psr\, with NGC6752.\ They questioned the arguments by
\cite{dpf+02}, which were based on the chance probability of finding a
pulsar in a GC observation and on the value of the
dispersion measure of \psr, which is very close to the value for the other four
pulsars known in NGC6752 at the time.

An updated ephemeris for \psr\, was reported by \cite{cpl+06}, who
obtained the first measurement of the pulsar's proper
motion based on the timing analysis of $\sim5$\,years of radio
observations with the Parkes `Murriyang' telescope, although the low precision prevented them from proving or disproving the  association of \psr\, with NGC6752. \cite{cpl+06} also searched for the signature of the Shapiro delay in the system but found no evidence of this phenomenon.

The first detection of the Shapiro delay in the \psr\, binary system
required $\sim$12\,yrs of timing observations with the
Parkes Murriyang radio telescope \citep{cbp+12}. From this detection, the timing analysis allowed the authors to measure the companion mass,
$M_{\rm C}=0.180\pm0.018M_\odot$, to put a lower limit on the orbital
inclination, $i\geq88^\circ$, and to derive, from these values and their measurement of the system's mass function, a conservative range for the pulsar mass, $1.1M_\odot\leq M_{\rm P}\leq1.5M_\odot$.

The GC NGC6752 has been widely observed in the last decade for a large number of scientific goals beyond pulsar astronomy. In particular, many positional and structural parameters have been updated since the \cite{cbp+12} paper.  In this work, we mention the latest total mass ($M$), parallax ($\varpi$), and core radius ($r_{\rm c}$) determinations: $M=(2.76\pm0.04)\times10^5M_\odot$ \citep{hbsb20}, $\varpi=0.251\pm0.010$\,mas \citep{vb21}, corresponding to a distance of $3.98\pm0.16$\,kpc, and $r_{\rm c}=0.13$\,arcmin \citep{hbsb20}. These values will be used throughout the paper.

In this paper, we report on $\sim$22\,years of observations of \psr\, obtained using the Parkes 64 m Murriyang telescope and, since 2020, 
the more sensitive MeerKAT radio telescope. The paper is organised as follows: In Sect. \ref{sec:obs} we describe the instruments and the data reduction techniques, while in Sect. \ref{sec:timing} we present our data analysis and discuss the properties of \psr\, that we can infer from our improved measurements. In Sect. \ref{sec:profile} we present a study of the pulsar profile and investigate the features of the interstellar medium (ISM) along the line of sight towards this object, while in Sect. \ref{sec:discussion} we discuss the implications of our results. In Sect. \ref{sec:summary} we briefly summarise our work.

\section{Observations and data analysis}
\label{sec:obs}

The data analysed in this work have been taken with the Parkes 64 m Murriyang and the MeerKAT radio telescopes over a total time span of $\sim$22 years. The data, observing systems used and the procedures for the extraction of the times of arrival (ToAs) are described below, while additional details are presented in Table\,\ref{table:observing_details}.

\renewcommand{\arraystretch}{1.2}
\begin{table*}[h]
 \caption[]{
 \label{table:observing_details}
 Observing logs and configurations used.}
 \centering
 \begin{threeparttable}
  \begin{tabular}{p{0.5in} l l l l l l c r }
   \hline
   \hline
   Telescope & Receiver & Backend & Centre   & Bandwidth$^\star$ & Number   & CD$^*$ & Time  & Number \\
             &          &         &Frequency &                  & of       &        & span  & of     \\
             &          &         &(MHz)     & (MHz)            & channels &        & (MJD) & TOAs   \\
   \hline
   \multirow{2}{*}{Parkes} & \multirow{2}{*}{H-OH} & AFB & 1390 & 256 & 512 & No & 53056$-$54223 & 76  \\
   &&PDFB4 & 1369 & 256 & 512 & No & 57472$-$57472 & 6  \\
      \hline
   & & AFB & 1390 & 256 & 512 & No & 51468$-$56205 & 954  \\
   Parkes&multibeam&PDFB3 & 1369 & 256 & 512-2048 & No &54833$-$56831 & 45 \\
   &&PDFB4 & 1369 & 256 & 1024 & No &54834$-$57425 & 127  \\
   \hline
   \multirow{4}{*}{MeerKAT} & \multirow{3}{*}{L-band} & PTUSE & 1283.58 & 775.75 & 928 & Yes & 58909$-$59350& 121\\
    &  & APSUSE & 1283.90 & 856 & 4096 & No & 59059 & 13  \\
    &  & PTUSE & 1283.90 & 856 & 4096 & Yes & 59389$-$59451 & 369  \\
    &  UHF-band & PTUSE & 815.73 & 544 & 1024 & Yes & 59040$-$59152 & 81  \\
   \hline
   \hline
  \end{tabular}
  \begin{tablenotes}
    $^\star$ Effective usable bandwidth. \\
    $^*$ Intra-channel coherent dedispersion.\\
   \end{tablenotes}
 \end{threeparttable}
\end{table*}

\subsection{Parkes observations and ToA extraction}
\label{subsec:pksobs}

Observations with the Parkes Murriyang radio telescope were carried out from September 1999 to March 2016, with the Multibeam and the H$-$OH receivers (depending on availability) at a central frequency of 1369\,MHz with a bandwidth of 256\,MHz. For most observations until September 2012, the signal was
1-bit digitised and recorded to magnetic tapes with the Analog
Filterbank (AFB) backend. In other cases, the signal was acquired in folding or search mode with the Pulsar Digital Filterbank 3 (PDFB3) and 4 (PDFB4). Technical details 
and references for this instrumentation can be found at the
website of the Parkes Radio Telescope\footnote{\url{https://www.parkes.atnf.csiro.au}}.

We folded the search mode data at the topocentric spin period of the pulsar with the {\tt dspsr} \citep{vb11} software package by using the latest published ephemeris from \citep{cbp+12}, with a typical integration length of 1 minute while
maintaining the same number of frequency channels of the raw data. Folding mode
data had been real time folded with the best
ephemeris available at each observation epoch. The same ephemeris was installed in these archives to maintain consistency.

We extracted the pulse ToAs using the Fourier domain Monte Carlo technique implemented in the routine
{\tt pat}, provided by the software suite {\tt psrchive} \citep{hvm04}
by convolving the observed pulse profiles with a high signal-to-noise
template obtained by summing in phase the brightest observed pulses
profiles. We built separate templates for each backend and observing mode
combination.
We also visually inspected each observed profile, and compared
them with the template, thus rejecting those ToAs whose corresponding
profile could not be evaluated as detected.
This scrutiny is necessary because of the strong signal scintillation
due to the ionised ISM, which causes significant changes in the pulsar's signal-to-noise ratio (S/N) at moderate dispersion measures up to a few tens of pc\,cm$^{-3}$. In the case of \psr, such changes are seen in
several observations in both time and frequency, not only between observations, but also within single observations.
For this reason we could not identify an optimal integration
length for obtaining ToAs of comparable S/N and
uncertainty, but we inspected each observation separately in order to
identify the extent in time and frequency along which pulses can be
considered detected. Once these time and frequency intervals were 
identified, we pursued the goal of obtaining the highest
reasonable number of ToAs, by partially summing the profiles in each data
file with respect to time and/or frequency. 

\subsection{MeerKAT observations and ToA extraction}
\label{subsec:mktobs}

The data from the MeerKAT telescope were obtained under two Large Survey Projects that include observations of GCs as part of their scientific goals. 

Most of these data were obtained from the MeerTime project \citep{bja+20} that focuses on timing known pulsars for a variety of scientific goals. The observations presented here were recorded in the context of either the globular cluster (GC) theme (e.g. \citealt{rgf+21}) or the relativistic binary (RelBin) theme \citep{ksv+21}. MeerTime observations used the Pulsar Timing User Supplied Equipment (PTUSE) backend. This backend acquires tied-array beam-formed voltages from the correlator-beam-former part of the MeerKAT observing system, and is capable of simultaneously recording coherently dedispersed full-Stokes parameters data, both in filterbank (search) mode and in folded archive mode. 
Additional data were collected under the TRAnsients and PUlsars with Meerkat (TRAPUM; \citealt{sk+16}) project, which is aimed at searching for new pulsars, including but not limited to searches of pulsars in GCs. The TRAPUM observations generally use the FilterBank User Supplied Equipment system to tile up to 768 tied-array beams on the sky, recording total intensity filterbank data per beam. A subset of these beams are then incoherently dedispersed and searched for pulsars with the Accelerated Pulsar Search User Supplied Equipment (APSUSE) backend. The results from these searches are or will be reported elsewhere. It is typical for observations of GCs to be also observed in parallel by PTUSE pointed at the cluster core, thereby providing complementary data with coherent dedispersion that is better for timing. However, these data were not useful for timing \psr\, given its large offset from the cluster centre. More recently, the PTUSE backend system has acquired the capability to record up to four tied-array beams on the sky, and this could be used in future to observe the source.

The data were acquired with two receivers: the L-band receiver that operates at a centre frequency of 1284.58\,MHz (with 1024 channels) or at 1283.90\,MHz (with 4096 channels) with a bandwidth of 856\,MHz, and the UHF  receiver that operates at a centre frequency of 815.73\,MHz with a bandwidth of 544\,MHz. The observations also included two 6-hour long observations carried out specifically around superior conjunction of \psr\, to maximise the sensitivity for the detection of the Shapiro delay.

The pulsar observing setup with MeerTime is explained in \cite{bja+20}. The \textsc{meerpipe} data reduction pipeline was used to process the raw data from the PTUSE machines. \textsc{meerpipe} performs excision of radio frequency interference using a modified version of \textsc{coastguard} \citep{LazarusEtAl2016}, followed by flux and polarisation calibration. The details on polarisation and flux calibration are outlined in \cite{sjk+21} and \cite{SpiewakEtAl2021}, respectively. 

The large bandwidth of MeerKAT receivers, combined with the higher sensitivity meant that we did not have to manually set the integration times on a per-observation basis as done for the Parkes data. 
For the MeerKAT ToA extraction, the calibrated data products from \textsc{meerpipe} were decimated to obtain total intensity profiles with eight channels across the band and 900 s integration lengths. Visual inspection confirmed that the pulses were clearly detected in all the eight frequency channels and all the integrations. Observations with high S/N were summed on a per backend/receiver basis to obtain good frequency resolved pulse profiles. We obtained 2D-analytical templates from these profiles by iteratively running the \textsc{paas} command from the \textsc{psrchive} software package for every channel. These templates were then used to obtain frequency resolved ToAs using the \textsc{pat} command.

\section{Timing analysis}
\label{sec:timing}

\begin{figure*}
    \centering
    \includegraphics[width=0.975\textwidth ]{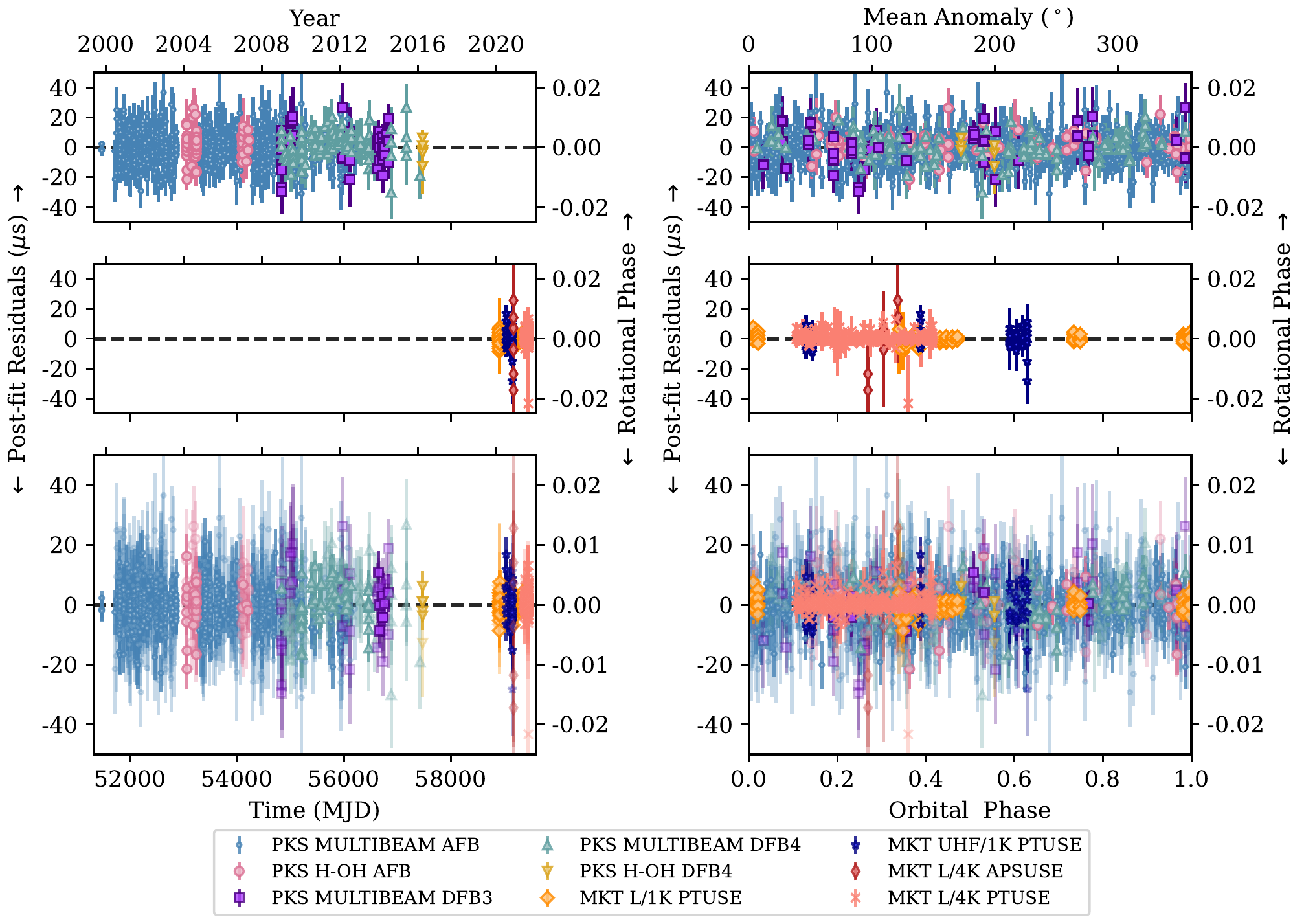}
    \caption{Post-fit timing residuals of \psr\, as a function of time and orbital phase using the ELL1 binary model. The first and second rows show the data from the 64 m Parkes Murriyang radio telescope and the MeerKAT radio telescope, respectively. The different receiver and backend combinations mentioned in Table \ref{table:observing_details} are denoted with different colours and symbols, respectively. The panels at the very bottom provide the combined dataset, with the ToAs that have uncertainties greater than 8$\mu$s made semi-transparent for clarity. }
    \label{fig:residuals}
\end{figure*}

\begin{figure*}
    \centering
    \includegraphics[width=\textwidth, trim={40 40 40 40} ]{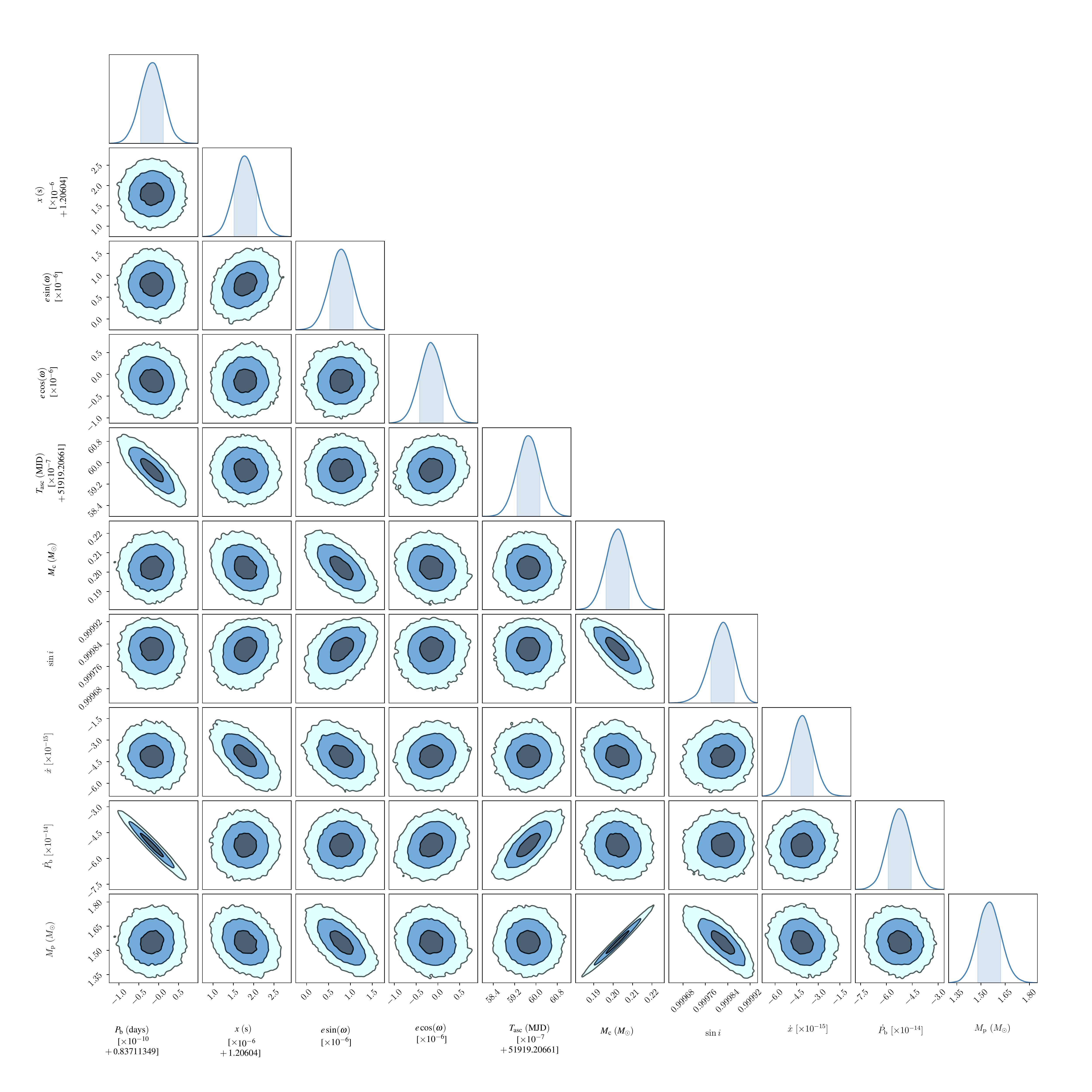}
    \caption{Corner plot showing the posterior distributions of the orbital, post-Keplerian parameters and the derived mass of \psr\, obtained from ELL1 binary model using \textsc{temponest}. The diagonal elements show the marginalised 1D histograms of the parameters while
    the off-diagonal elements show the correlation between the parameters and are marked by contours that define the 39\%, 86\%, and 98\% C.L. The shaded region in the 1D histograms indicates the nominal 68.27\% C.L. as noted in Table \ref{tab:binary}}
    \label{fig:corner_orbit}
\end{figure*}

We used the \textsc{tempo2} \citep{HobbsEtAl2006} timing software with the DE436 Solar System ephemeris published by the Jet Propulsion Laboratory\footnote{\url{https://www.jpl.nasa.gov}} for the initial timing analysis, and the \textsc{temponest} \citep{LentatiEtAl2014} parameter estimation plug-in to \textsc{tempo2} to perform non-linear fits of the timing model to the data. Tables\,\ref{tab:timing_params} and \ref{tab:binary} present the timing results, including the values of derived parameters of interests for this work and, for any mentioned parameter, the representing symbol. The timing residuals with respect to both epoch and orbital phase are displayed in Fig.\,\ref{fig:residuals}, while the corner plot of the non-linear fit is displayed in Fig.\,\ref{fig:corner_orbit}.

\begin{table*}
\caption{Timing parameters for \psr, obtained from the {\sc tempo2} timing package using the ELL1 binary model. The values in the parentheses indicate nominal 1$\sigma$ symmetric uncertainties on the last digit of the value. Asymmetric uncertainties are explicitly provided. 
}
\centering 
\begin{tabular} {l c}
\hline
\hline
\multicolumn{2}{c}{Adopted observation and data reduction parameters}\\
\hline
Solar System ephemeris\dotfill & DE436 \\
Timescale \dotfill & TCB \\
Reference epoch for spin frequency, position and DM (MJD)\dotfill & 59451\\
Solar wind electron number density, $n_{0}$ (cm$^{-3}$)\dotfill & 4 \\
\hline
\multicolumn{2}{c}{Spin and astrometric parameters}\\
\hline
Right ascension, $\alpha$ (J2000, h:m:s)\dotfill & 19:11:42.74680(2) \\
Declination, $\delta$ (J2000, d:m:s)\dotfill & $-$59:58:26.9850(3) \\
Proper motion in $\alpha$, $\mu_{\alpha}\cos\delta$ (mas\,yr$^{-1}$)\dotfill & $-3.23(2)$  \\
Proper motion in $\delta$, $\mu_{\delta}$ (mas\,yr$^{-1}$)\dotfill  & $-3.93(3)$ \\
Spin frequency, $\nu$ (Hz)\dotfill & $306.16744090851(7)$ \\
First derivative of spin frequency, $\dot{\nu}$ ($10^{-16}$\,Hz\,s$^{-1}$)\dotfill &$ -2.7407^{+0.0007}_{-0.0005}$  \\
Second derivative of spin frequency, $\ddot{\nu}$,($10^{-27}$\,Hz\,s$^{-2}$)\dotfill &$ 5.8(2)$  \\
Dispersion measure, DM (cm$^{-3}$\,pc)\dotfill  &  33.704(2) \\
First Derivative of DM, DM1 ($10^{-4}$\,cm$^{-3}$\,pc\,yr$^{-1}$)\dotfill  & $-12^{+7}_{-6}$ \\
Second Derivative of DM, DM2 ($10^{-5}$\,cm$^{-3}$\,pc\,yr$^{-2}$)\dotfill  & $-13^{+7}_{-6}$ \\
Rotation measure, RM (rad\,m$^{-2}$)\dotfill  & $43.0(2)^{\textit{a}}$ \\

\hline
\multicolumn{2}{c}{Derived parameters}\\
\hline
Galactic longitude, $l_{\rm GAL}$ ($^{\circ}$)\dotfill  & 336.525  \\
Galactic latitude, $b_{\rm GAL}$ ($^{\circ}$)\dotfill  & $-$25.730 \\
DM-derived distance (NE2001), $D_{\rm NE}$ (kpc)\dotfill & 1.153  \\
DM-derived distance (YMW16), $D_{\rm YMW}$ (kpc)\dotfill & 1.685  \\
Galactic height, $z$ (kpc)\dotfill & $-1.79^{\textit{b}}$  \\
Position angle of proper motion, J2000, $\Theta_{\mu}$ ($^{\circ}$) \dotfill &  219.4(3) \\
Position angle of proper motion, Galactic, $\Theta_{\mu}^{\rm Gal}$ ($^{\circ}$) \dotfill & 299.5(3) \\
Total proper motion, $\mu$ (mas\,yr$^{-1}$)\dotfill  & 5.087(26) \\
Heliocentric transverse velocity, $V_{\text{T}}$ (km\,s$^{-1}$)\dotfill & $95(4)^{\textit{b}}$  \\
Spin period, $P_{0}$ (ms)\dotfill & 3.2661866214323(8) \\
Spin period derivative, $\dot{P}_{0}$ ($10^{-21}$\,s\,s$^{-1}$)\dotfill & 2.96793(64) \\
Contribution to $\dot{P}$ from the variation of the Doppler shift, $\dot{P}_{\text{k}}$ ($10^{-21}$\,s\,s$^{-1}$)\dotfill & $-2.05(30)^{\textit{c}}$\\
Intrinsic spin period derivative, $\dot{P}_{\rm 0,intr}$ ($10^{-21}$\,s\,s$^{-1}$)\dotfill & 5.02(30) \\
Surface magnetic field strength, $B_{\text{surf}}$ ($10^{8}$\,G)\dotfill  & 1.29(4) \\
Characteristic age, $\tau$ (Gyr)\dotfill & 10.3(6) \\
Spin-down power, $\dot{E}_{\rm rot}$ ($10^{33}$\,erg\,s$^{-1}$)\dotfill & $-$5.7(3)  \\
\hline
\multicolumn{2}{c}{Fixed parameters}\\
Parallax, $\varpi$ (mas)\dotfill  & 0.251(10)$^{\textit{b}}$\\ 
\hline
\hline
\multicolumn{2}{l}{$^{\textit{a}}$ Obtained using \textsc{rmfit} program in the \textsc{psrchive} software package.}\\
\multicolumn{2}{l}{$^{\textit{b}}$ Assuming that the distance is the same as NGC6752.}\\
\multicolumn{2}{l}{$^{\textit{c}}$ See Eq.~\ref{eq:pdiscr}.}\\
\label{tab:timing_params}
\end{tabular}
\end{table*}

\begin{table*}
\caption{Timing parameters (continuation of Table \ref{tab:timing_params}) for \psr, obtained from the {\sc tempo2} timing package using the ELL1 (second column) and ELL1H (third column) binary model. The values in the parentheses indicate nominal 1$\sigma$ symmetric uncertainties on the last digit of the value. Asymmetric uncertainties are explicitly provided. The values in curly brackets indicate derived quantities from the measurement. The values of $x$ and $\epsilon_1$ for the ELL1H model are provided after subtracting the contributions from the first and second harmonics of the Shapiro delay (see Eqns. 16 and 17 in \citealt{FreireAndWex2010} for details).
}
\centering 
\begin{tabular} {l c c }
\hline
\hline
Binary model\dotfill & ELL1 & ELL1H  \\[0.5ex] 
Number of ToAs \dotfill & 1788 & 1788 \\
Weighted rms of ToA residuals ($\mu s$) \dotfill & 2.17  & 2.13\\
\hline
\multicolumn{3}{c}{Orbital parameters}\\[0.5ex]
\hline
Orbital period, $P_{\text{B}}$ (days)\dotfill & 0.837113489987(3) & 0.837113489970(4)  \\[0.5ex]
Projected semi-major axis of the pulsar orbit, $x$ (s)\dotfill & 1.20604175(3)& 1.20604176(3)\\[0.5ex]
Epoch of ascending node passage, $T_{\rm ASC}$ (MJD)\dotfill &51919.206615968(5) & 51919.206615991(5)  \\[0.5ex]
$1^{\rm st}$ Laplace-Lagrange parameter, $\epsilon_1 =  e \sin \omega (10^{-7})$ \dotfill & 8.0(3) & 8.5(3) \\[0.5ex]
$2^{\rm nd}$ Laplace-Lagrange parameter, $\epsilon_2 =  e \cos \omega (10^{-7})$\dotfill & $-$1.8(3)  & $-$1.7(3) \\[0.5ex]
Rate of change of projected semi-major axis, $\dot{x}$ ($10^{-16} \rm ls\,s^{-1}$)\dotfill & $ -40.7^{+7.3}_{-8.2} $ & $-40.3^{+8.0}_{-9.3} $ \\[0.5ex]
Orbital period derivative, $\dot{P_{\text{B}}}$ ($10^{-15}$\,s\,s$^{-1}$)\dotfill & $ -53.0^{+7.4}_{-6.0} $ & $-49.3^{+7.2}_{-8.0}$ \\[0.5ex]
Range of Shapiro delay, $r$ ($T_\odot$)\dotfill &       $0.202(6) $&  \{$0.201(7) $\} \\[0.5ex]
Shape of Shapiro delay, $s$\dotfill & $0.999823(4)$ &  \{$0.999834^{+0.00004}_{-0.00005}$\} \\[0.5ex]
Orthometric amplitude of Shapiro delay, $h_{3}$ ($\mu$s)\dotfill &  \{$0.94(2)$\} & $0.938(3)$ \\[0.5ex]
Orthometric ratio of Shapiro delay, $\varsigma$\dotfill & \{$0.981(2)$\} & $0.982(3)$ \\[0.5ex]
\hline
\multicolumn{3}{c}{Noise parameters}\\
\hline
Red noise power-law amplitude, $A_{red}$\dotfill & $-14.33^{+0.68}_{-0.85}$ & $-15.41^{+0.87}_{-0.48}$ \\
Red noise power-law spectral index, $\alpha_{red}$\dotfill & $6.4(1)$ & $7.95^{+0.96}_{-1.39}$ \\
DM noise power-law amplitude, $A_{DM}$\dotfill & $-10.59^{+0.04}_{-0.03}$ & $-10.59(4)$ \\
DM noise power-law spectral index, $\alpha_{DM}$\dotfill & $1.75^{+0.17}_{-0.14}$ & $1.79^{+0.16}_{-0.14}$ \\
\hline
\hline
\multicolumn{3}{c}{Derived masses and inclination }\\
\hline
Mass function, $f$ (M$_{\odot}$)\dotfill & $0.0026878255(2)$ & $0.0026878256(2)$\\[0.5ex]
Orbital inclination, $i$ ($\deg$)\dotfill & $88.90^{+0.15}_{-0.14}$ & $88.9(2)$\\[0.5ex]
Companion mass, $M_{\text{C}}$ (M$_{\odot}$)\dotfill  &$0.202(6) $&  $0.201(7) $ \\[0.5ex]
Pulsar mass, $M_{\text{P}}$ (M$_{\odot}$)\dotfill  & $1.556^{+0.067}_{-0.076}$ & $1.541^{+0.080}_{-0.088}$ \\[0.5ex]
\hline
\hline

\label{tab:binary}
\end{tabular}
\end{table*}

The fitted timing model includes rotational (spin frequency and derivatives), astrometric (position and proper motion), ISM (dispersion measure and derivatives) and orbital (Keplerian and post-Keplerian) parameters. We also included timing `jumps' to account for arbitrary time offsets between different telescopes/receivers/backends/frequencies. We described the  pulsar's orbital motion with the ELL1 \citep{lcw01}, which is particularly suitable for treating the dynamics of binaries with very low orbital eccentricities ($e$). For such binaries, the location of the periastron (parameterised by its longitude relative to the ascending node, $\omega$) cannot be precisely determined, which implies that the time of passage through periastron ($T_0$) cannot be determined precisely either; generally these two parameters have a very large correlation. The ELL1 model avoids this by using instead the time of ascending node ($T_{\rm asc}$), which can be determined very precisely even for circular orbits, and the Laplace-Lagrange parameters, $\epsilon_1 \, \equiv e \sin \omega$ and $\epsilon_2 \, \equiv e \cos \omega$.

An initial fit of the model to the data was performed using the \textsc{tempo2} timing software. We thus obtained reasonable prior probabilities of the timing parameters, used as input to \textsc{temponest} to perform non-linear fits of the timing model.

We also included stochastic parameters to characterise the noise in the data, namely: a white noise model (WN) with parameters EFAC that scale and add to the ToA uncertainties to compensate for stochastic variations of the pulse profile and instrumental noise; a red noise power law model (RN), parameterised by its amplitude, $A_{\rm red}$, and spectral index, $\alpha_{\rm red}$, to remove the secular timing noise that arises from intrinsic emission irregularities; and a dispersion measure (DM) power law model (DMN), parameterised by its amplitude, $A_{\rm DM}$, and spectral index, $\alpha_{\rm DM}$, that describe the temporal evolution of DM as a chromatic red noise. Further details about the noise models can be found in \cite{LentatiEtAl2014}. We performed 4 different combinations of noise model fits to our data that included (1) WN only; (2) WN+DMN; (3) WN+RN; and (4) WN+DMN+RN. We performed these fits thrice, first with no parallax ($\varpi_0$) in our timing model, second with parallax set to the best value of the cluster parallax ($\varpi_{\rm GC}$) as obtained from \citet{vb21}, and finally with parallax as a free model-parameter ($\varpi_{\rm f}$) for a total of 12 different non-linear fits of the timing model to the data. 

We computed the Bayes factors (BFs) between all possible combinations of the above mentioned model combinations: the noise in the data were best were best characterised by WN+DMN+RN  (with BF $\sim600$ compared to WN only, $\sim470$ compared to WN+RN and $\sim6$ compared to WN+DMN). The BFs 
did not provide any appreciable difference between the models containing $\varpi_0,\varpi_{\rm GC}$ and $\varpi_{\rm f}$ (with BF $\leq 2$)
Therefore, we chose the model with WN, DMN, RN, and $\varpi_{\rm GC}$ as the best model to describe the dataset given our demonstration, reported in Sect. \ref{subsec:PSRAassoc}, that \psr\, is associated with the GC NGC6752.

As a consistency check of the estimated the binary parameters, we repeated the analysis described above with the best noise/parallax combination by using the ELL1H binary model. This model is identical to the ELL1 model in all other aspects except for the parameterisation of the Shapiro delay \citep{FreireAndWex2010}. It accounts for the Shapiro delay by measuring the amplitude of its third harmonic (orthometric amplitude, $h_{\rm 3}$) and the ratio of the amplitudes of successive harmonics (orthometric ratio, $\varsigma$) . Table\,\ref{tab:binary} and Fig.\,\ref{fig:corner_orbit}, where the results obtained with both models are simultaneously presented/displayed for immediate comparison, show an excellent agreement between the results obtained from the two binary models. In the remainder of the paper we comment on and make use of the values obtained with the ELL1 model.

\subsection{Mass measurements}
\label{subsec:shapiro}

The dense observations around superior conjunction have allowed
precise measurements of two post-Keplerian parameters, the range ($r$) and shape ($s$) of Shapiro delay.
Assuming general relativity (GR), one can relate these measurements to the system parameters as
\begin{alignat}{2}
r &= T_{\odot}\, M_{\rm C} \label{eq:shr}\\
s &\equiv \sin i, \label{eq:shs}
\end{alignat}

\noindent
 where $T_\odot \equiv ({\cal G} {\cal M})_{\odot}^{\rm N} / c^3 = 4.9254909476412669... \mu$s is an exact number derived from the nominal solar mass parameter (see \citealt{2016AJ....152...41P}).
Once Eqs. \ref{eq:shr} and \ref{eq:shs} are
combined with the mass function,

\begin{equation}
    f(M) \equiv \frac{4 \pi^2}{T_{\odot}}\frac{x^3}{P_{\rm B}^2} = \frac{(M_{\rm C} \sin i)^3}{(M_{\rm P}+M_{\rm C})^2} = 0.002687826(2) \rm M_{\odot},  
\end{equation}

\noindent
we can determine $M_{\rm P}$, $M_{\rm C}$ and $\sin i$. From the marginalised 1D probabilities displayed in Fig.\,\ref{fig:corner_orbit} we obtain
$M_{\rm C} = 0.202\pm0.006M_{\odot}$ and $M_{\rm P} = 1.556^{+0.067}_{-0.076} M_{\odot}$. The measurement of $M_{\rm C}$ is in full agreement with all previous limits or estimates (\citealt{fpsd03}, \citealt{cfpd06}, \citealt{bkkv06}, \citealt{cbp+12}), while the inferred value for $M_{\rm P}$ is consistent with the one obtained by \cite{cbp+12} from their low-precision detection of the Shapiro delay. From the value of $\sin i$, $i$ is either $88.90^{+0.15}_{-0.14}\,\deg$ or $91.10^{+0.14}_{-0.15}\,\deg$, which makes this system one of the most edge-on binary pulsar systems currently known. Under several assumptions, we can break  the degeneracy between the two values of $i$ using polarisation data as explained in Sect. \ref{sec:rvm}. 

Given that we now have a precise estimate of the proper motion and the distance (given by the cluster distance; see Sect. \ref{subsec:PSRAassoc}), analysis of the pulsar's scintillation over orbital and yearly timescales could potentially provide an independent estimate of the sense of $i$ (e.g. \citealt{ReardonEtAl2018,ReardonEtAl2020}), although the current frequency resolution is too coarse to resolve the scintillation structure.

\subsection{The intrinsic spin-down of \psr}
\label{subsec:truepdot}

The measurement of the Shapiro delay in Sect. \ref{subsec:shapiro} allows
the calculation of the amplitude of all other post-Keplerian parameters assuming GR adequately describes the dynamics of the system. In particular the  predicted rate of the orbital decay,  $\dot{P}_{\rm B,GR} = (-7.632\pm0.024)\times10^{-14}$\,s\,s$^{-1}$ is in clear disagreement with the observed value $\dot{P}_{\rm B,obs} \equiv \dot{P}_{\rm B} = (-5.30_{-0.60}^{+0.74})\times10^{-14}$\,s\,s$^{-1}$. By assuming that
the orbital decay is mainly responsible for the intrinsic variation
of the orbital period, that is to say, that all other phenomena give a negligible
contribution, we can say $\dot{P}_{\rm B,GR} = \dot{P}_{\rm B,intr}$.
The discrepancy can be ascribed to the acceleration of the source with respect to the Solar System barycentre, according to
\begin{equation}
\left(\frac{\dot{P}_{\rm B}}{P_{\rm B}}\right)_{\rm obs} \,  = \,
\left(\frac{\dot{P}_{\rm B}}{P_{\rm B}}\right)_{\rm intr}\, + \, \frac{\mathcal{A}_{\rm l}}{c}, 
\label{eq:pdiscr}
\end{equation}
where $\mathcal{A}_{\rm l}$ is the component along the line of sight of the sum of all accelerations, both real and apparent, acting on
the binary system, this will be discussed in detail in Sect.~\ref{sec:discussion}.  A similar relation holds true for all the periodic physical quantities associated with the source. Therefore,  taking the equivalent expression for the spin period and subtracting Eq.~\ref{eq:pdiscr} from it, we can determine the only unknown term:

\begin{equation}
\left(\frac{\dot{P}_{\rm 0}}{P_{\rm 0}}\right)_{\rm intr}\, = \, 
\left(\frac{\dot{P}_{\rm 0}}{P_{\rm 0}}\right)_{\rm obs}\, - \,
\left(\frac{\dot{P}_{\rm B}}{P_{\rm B}}\right)_{\rm obs}\, + \,
\left(\frac{\dot{P}_{\rm B}}{P_{\rm B}}\right)_{\rm intr}.
\label{eq:accels}
\end{equation}
From this, we obtain $\dot{P}_{\rm 0,intr}/P_{\rm 0}=(9.09\pm0.20)\times10^{-19}$\,s$^{-1}$, and hence $\dot{P}_{\rm 0,intr}=(+5.02\pm0.30)\times10^{-21}$\,s\,s$^{-1}$. This value is well within the observed $\dot{P}$ of the Galactic millisecond pulsar (MSP) population, although towards the lower end of the range: only 40 out of 206 objects\footnote{Those include all the MSPs with a measured non negative value of the first derivative of the spin period in the database version 1.67 of the ATNF pulsar catalogue {\tt psrcat}, available at \url{https://www.atnf.csiro.au/research/pulsar/psrcat/download.html}} show $\dot{P}<\dot{P}_{\rm 0}.$ The measurement of $\dot{P}_{\rm 0,intr}$ allows us to infer the characteristic age $\tau=10.3\pm0.6$\,Gyr, the spin-down luminosity $\dot{E}_{\rm rot}=(-5.68\pm0.34)\times10^{33}$\,erg\,s$^{-1}$ and the surface magnetic field $B_{\rm surf}=(1.30\pm0.04)\times10^8$\,G for \psr.

\subsection{Rate of change of the length of the projected semi-major axis}
\label{sec:xdot}
The timing analysis provides a tantalising 5$\sigma$ measurement of the secular evolution of the length of the projected semi-major axis of the orbit of \psr\,, $\dot{x}_{\rm obs}=-4.1^{+0.7}_{-0.8} \times 10^{-15}$. This can, in principle, arise due to a number of physical and geometric contributions, which can be decomposed as  
\begin{equation}
\dot{x}_{\mathrm{obs}} = \dot{x}_{\rm PM} + \dot{x}_{\rm \dot{D}} + \dot{x}_{\rm GW} + \dot{x}_{\dot{\rm M}} + \dot{x}_{\rm 3^{rd}} +  \dot{x}_{\dot{\epsilon}_{\rm A}} + \dot{x}^{\rm P}_{\rm SO} + \dot{x}^{\rm C}_{\rm SO}.
\end{equation}
 We now describe each term and estimate its magnitude:
The first term, $\dot{x}_{\rm PM}$ is caused by the proper motion of the system. It is given by \citep{Kopeikin1996}
\begin{equation}
\label{eqn:xdot_pm}
\dot{x}_{\rm PM} \leq1.54 \times 10^{-16}  x \cot i \left( \frac{\mu}{\rm mas~yr^{-1}} \right).
\end{equation}
Since both the total proper motion $\mu$ and $i$ are well constrained (see Table \ref{tab:timing_params}), this yields a contribution of order $10^{-19}$,  which is four orders of magnitude below $\dot{x}_{\rm obs}$; therefore, this term is negligible.

The second term, $\dot{x}_{\rm \dot{D}}$, is due to the changing radial Doppler shift, $\cal{A}_{\rm l}$, which includes the pulsar's acceleration in the gravitational field of the cluster. The third term, $ \dot{x}_{\rm GW}$, is caused by the shrinkage of the orbit caused by gravitational wave emission. The fourth term, $\dot{x}_{\dot{\rm M}}$, is caused by possible mass loss in the system. These three effects contribute to both $\dot{P}_{\rm B}$ and $\dot{x}$. Since we have a measurement of $\dot{P}_{\rm B}$, the contributions from these effects to both the measurements can be related, to within the same order of magnitude, as (see e.g. \citealt{DamourTaylor1992})
\begin{equation}
    \label{eqn:xdot_pbdot}
    \left(\frac{\dot{x}}{x}\right) \sim  \left(\frac{\dot{P}_{\rm B}}{P_{\rm B}}\right).
\end{equation}
From this, we obtain a total corresponding contribution from these terms to $\dot{x}_{\rm obs}$ of the order of $10^{-19}$. Hence, these contributions are also negligible.

The fifth term, $\dot{x}_{\rm 3^{rd}}$ is caused by the presence of a hypothetical third body in the system. However, there are two arguments that make this unlikely.
 First, the presence of a third body would induce significant drifts in the timing of the pulsar that usually require the addition of several higher-order spin derivatives. We find no evidence for spin derivatives higher than $\ddot{\nu}$; the magnitude of $\dot{\nu}$ and $\ddot{\nu}$ are normal for pulsars in GCs that are affected by cluster acceleration. Second, the extremely circular nature of the binary with eccentricity of the order of $10^{-7}$ is further strong evidence against any acceleration towards a third body, which would naturally induce eccentricity in binary systems with high mass ratios (e.g. \citealt{Wolszczan1991}). 

This leaves two residual contributors as the only possible explanation for $\dot{x}_{\rm obs}$: $\dot{x}_{\dot{\epsilon}_{\rm A}}$ is the secular change in the aberration of the pulsar beam due to geodetic precession (where $\epsilon_{\rm A}$ is the first aberration parameter) and $\dot{x}^{\rm P}_{\rm SO}$ and $ \dot{x}^{\rm C}_{\rm SO}$ result from the spin-orbit coupling from the fast spin of the pulsar and the companion \citep{DamourTaylor1992, Lorimer&Kramer2005}. Each of the last two terms results from the sum of two effects, the first of them is the Newtonian precession caused by its quadrupole moment, the second is the relativistic Lense-Thirring (LT) effect.
We estimate these contributions in the Appendix, finding that $\dot{x}_{\rm obs}$ can be predominantly ascribed to the quadrupolar moment of a white dwarf (WD)  companion rotating with a period of a few hours.

This explanation requires the spin of the WD to be misaligned with the orbital angular momentum; this is necessary in order for the latter vector to precess around the total angular momentum vector. The problem with this explanation is that such a misalignment is certainly unexpected from the evolution of MSP-HeWD systems, where the transfer of angular momentum should result in the pulsar having an angular momentum that is exactly parallel to the orbital angular momentum. In Sect.~\ref{sec:discussion} we discuss the implications in more detail.

\begin{figure*}
    \centering
    \includegraphics[width=\textwidth]{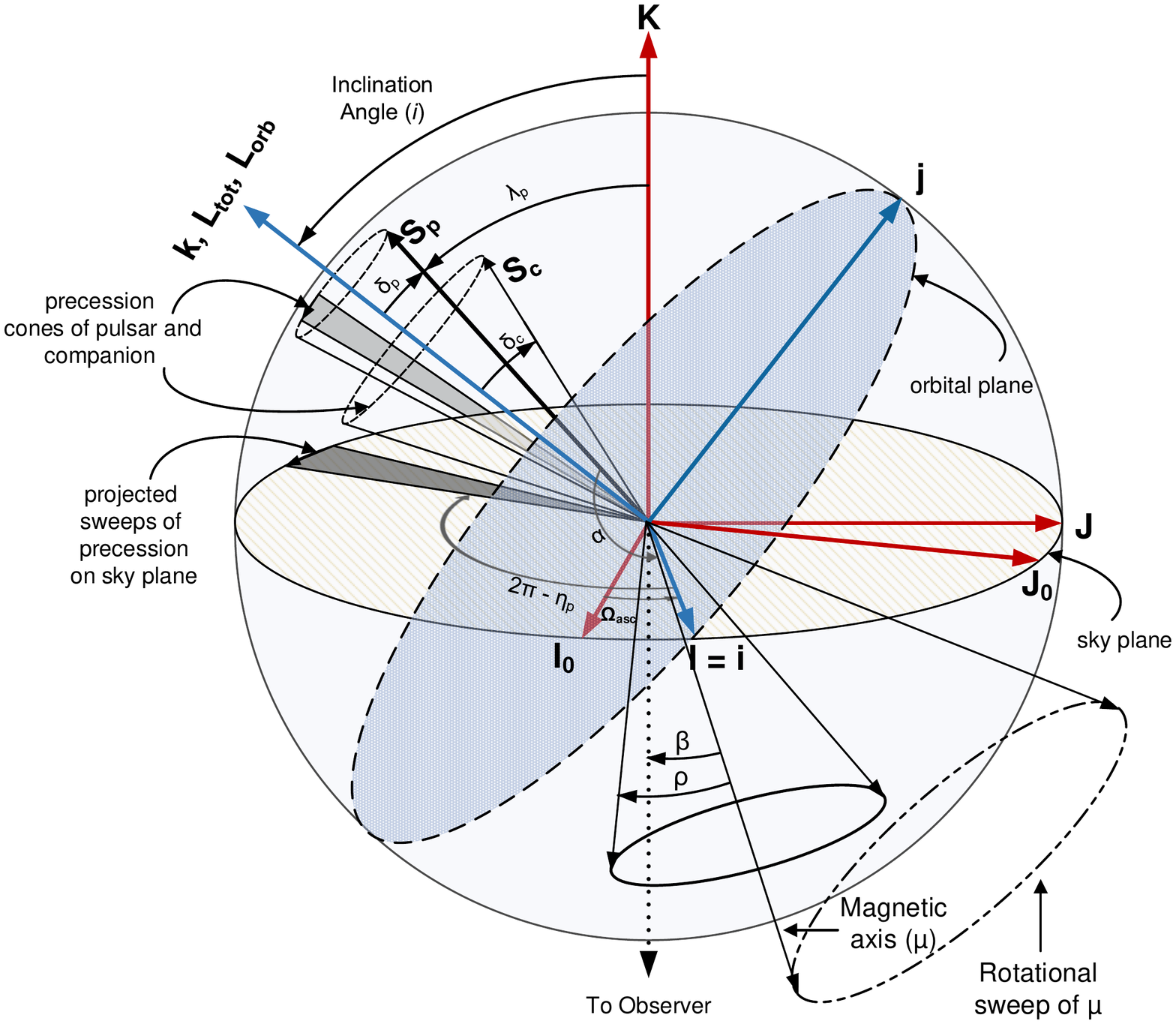}
    \caption{Definition of angles and vectors used, following the DT92 convention \citep{DamourTaylor1992}. All markers in boldface denote vector quantities. Let {\bf K} be the line-of-sight vector defined to be from the observer to the pulsar. The plane perpendicular to this vector forms the sky plane, defined by unit vectors $\mathbf{I_0}$ and $\mathbf{J_0}$. The orbital angular momentum ($\mathbf{L_{\rm orb}}$) is in the direction of {\bf k} and is inclined from {\bf K} by the orbital inclination angle, $i$. The plane perpendicular to this, defined by unit vectors {\bf i} and {\bf j}, is the orbital plane. $\Omega_{\rm asc}$ is the angle of rotation of the sky plane with respect to the orbital plane when the pulsar passes through the ascending node.  
    The spin of the pulsar and the companion, defined by vectors $\mathbf{S_{\rm p}}$ and $\mathbf{S_{\rm c}}$, respectively, are misaligned with $\mathbf{L_{\rm orb}}$ by the spin-misalignment angles $\delta_{\rm p}$ and  $\delta_{\rm c}$, respectively. $\mathbf{S_{\rm p}}$ and $\mathbf{S_{\rm c}}$ and $\mathbf{L_{\rm orb}}$ all precess around the total angular momentum of the system, $\mathbf{L_{\rm tot}}$, which is the vector sum of all the individual angular momenta. However, since the magnitude of $\mathbf{L_{\rm orb}}$ is orders of magnitude more than $\mathbf{S_{\rm p}}$ and $\mathbf{S_{\rm c}}$, $\mathbf{L_{\rm tot} = L_{\rm orb}}$ is assumed in the figure for clarity. Hence, $\mathbf{S_{\rm p}}$ and $\mathbf{S_{\rm c}}$ precess around {\bf k}, forming precession cones as shown. The projection of $\mathbf{S_{\rm p}}$ and $\mathbf{S_{\rm c}}$ on to the sky plane subtends an angle $\eta_{\rm p}$  and $\eta_{\rm c}$,  respectively -- for clarity, only the pulsar's complementary angle is shown. The pulsar's magnetic axis ($\mu$) subtends an angle $\alpha$ with respect to $\mathbf{S_{\rm p}}$. $\delta$ denotes the opening angle of the emission cone of the pulsar. As the pulsar rotates, $\mu$ sweeps across the sky; the angle suspended by $\mu$ during its closest approach with respect to our line of sight is the impact angle $\beta$. $\lambda_{\rm p}$ is the angle between {\bf K} and $\mathbf{S_{\rm p}}$. Hence, by definition, $\lambda_{\rm p} \equiv 180 - \zeta \equiv 180 - (\alpha + \beta) $, which are the angles used by the RVM.}
    \label{fig:geometry}
\end{figure*}

\section{Profile analysis}
\label{sec:profile}

\begin{figure*}
    \centering
\includegraphics[width=0.99\textwidth, trim={25 25 25 25} ]{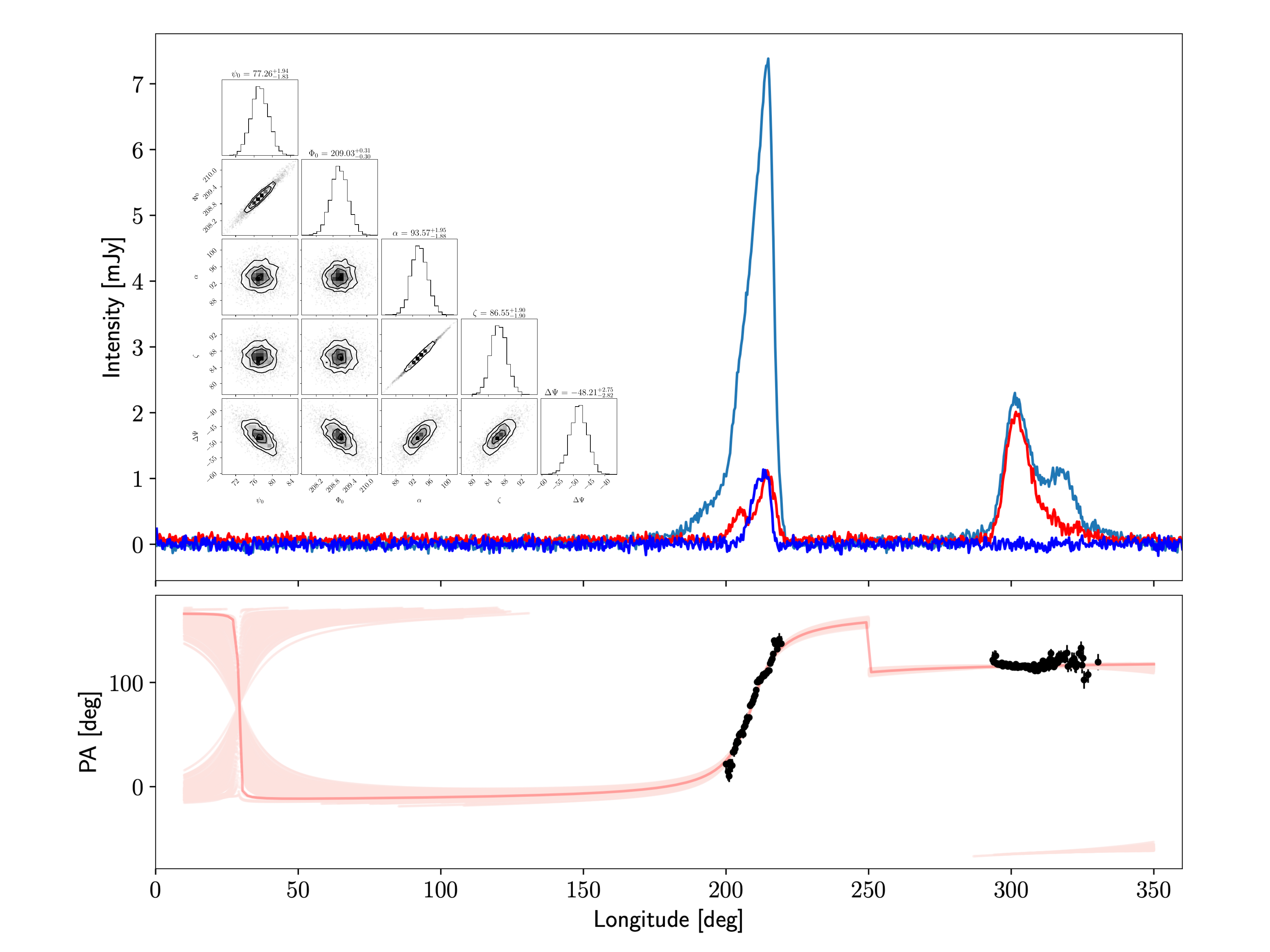}
    \caption{Flux and polarisation calibrated profile of \psr\, obtained using the L-band receiver of MeerKAT. {\em Top panel:} Total intensity, linearly polarised intensity, and circularly polarised intensity shown with the light blue, red, and dark blue lines, respectively. {\rm Bottom panel:} Measured position angle of the linearly polarised component as a function of pulse longitude. The red lines denote different realisations of the modified RVM fit that accounts for a vertical shift of position angle points for the post-cursor, as suggested by \cite[see the detailed discussion in text]{Dyks2019}, with the darkest line denoting the fit based on the maximum likelihood values. The inset in the top panel shows a corner plot of the posterior distributions of the RVM model parameters, with the off-diagonal elements representing the correlations between parameters and the diagonal elements denoting the marginalised histograms.}
    \label{fig:profile}
\end{figure*}

Figure\,\ref{fig:profile} shows the summed profile of all MeerKAT L-band observations taken  with 1K mode (that records 1024 frequency channels across the band; see \citealt{BailesEtAl2020})
At first glance, the two distinct components of the pulse profile look like main pulse and inter-pulse emission from opposite poles. However, for reasons that we will get to later, we generalise the terminology and name the brighter pulse as the `main pulse' and the other one as the `post-cursor'.

\subsection{Integrated total-intensity profile}

We measured the mean flux density in the L band using MeerKAT data to be 0.304(7) mJy. The polarisation fraction for both the main pulse and the post-cursor is at the few percent level, with the main pulse showing significant circular polarisation that is absent in the post-cursor. We also measured the rotation measure, $\rm RM \, = \, 43.0(2)\, \rm rad~m^{-2}$, by using the \textsc{rmfit} program of the \textsc{psrchive} software package.

\subsection{Scattering}

The steep drop in the main pulse of the integrated profile shape (Fig.\,\ref{fig:profile}) suggests that the observation is unlikely to suffer from measurable pulse broadening due to multi-path propagation in the ISM. We tested this hypothesis by modelling the integrated profile as an intrinsic five-component Gaussian shape convolved with an ISM transfer function described by an exponential decay ($e^{-t/\tau_{\rm scat}}$) and characteristic scattering timescale, $\tau_{\rm scat}$, as in, for example, \citet{Williamson1972}. The resulting fits across four frequency channels provide scatter broadening, $\tau_{\rm scat}$, values with large error bars and are consistent with zero-phase bins, such that there is no  measurable frequency evolution of $\tau_{\rm scat}$. This means that we cannot reliably estimate the power law index $\alpha_{\rm scat}$, which is used to describe the scattering as a function of frequency using $\tau_{\rm scat} \propto \nu^{-\alpha_{\rm scat}}$; this value is typically 4 or 4.4 for simple scattering models within the ionised ISM. We note that the $\tau_{\rm scat}$ values we have obtained are highly correlated (>0.98) with other parameters, especially the centroid values of the five Gaussian components. 

We also fitted the post-cursor component independently with a scattering model consisting of a two component intrinsic Gaussian convolved with the ISM transfer function as above, across four frequency channels. In this case we find apparently significant values on $\tau_{\rm scat}$, with, for example, $\tau_{\rm scat} = 0.165 \pm 0.020$\,ms at 1.4 GHz, but again note high correlations (>0.9) between $\tau$ values and Gaussian components (widths, centroids and amplitudes). We find $\alpha_{\rm scat} = -0.1 \pm 0.2$; once more, there is no clear frequency dependence of $\tau_{\rm scat}$, which again indicates no real detection of scattering. Keeping $\tau_{\rm scat}$ fixed at these values per channel and redoing the five component profile fit leads to a model that clearly overestimates the scattering on the steep trailing edge of the main pulse such that the best-fit obtained $\tau_{\rm scat}$ values from the post-cursor
component cannot describe the scattering of the full profile shape.  

Lastly, we conducted similar scattering tests using the UHF observations, where scattering is expected to be enhanced. Since our S/N for these data is significantly lower than in the L band, we only considered a frequency-averaged profile shape. An estimate on the upper limit of $\tau_{\rm scat}$ is obtained by keeping the centroids of the intrinsic Gaussian components fixed at its best-fit values from the highest L-band frequency channel analysed previously. In doing so, we obtain a $\tau_{\rm scat}$ value of $1.8 \pm 0.6\,\mu$s, which in phase bins gives $\tau_{\rm scat} = 0.58 \pm 0.19$ bins. We conclude that we do not find evidence for scattering in the temporal domain when analysing the full profile shape; {this effect is, as expected, too small to be measurable}. 

\subsection{Pulsar and orbital geometry from pulse structure data}
\label{sec:rvm}

The variation in the position angle ($\psi$) of linearly polarisation of a pulsar, if it arises solely due to geometric reasons, can be
described by the rotating vector model (RVM; \citealt{Radhakrishnan&Cooke}). The RVM describes $\psi$ as a function of the pulse phase, $\Phi$, depending on the magnetic inclination angle, $\alpha$ and the viewing angle, $\zeta$, which is the angle between the line-of-sight vector and the pulsar's spin and can be written as
\begin{equation}
\label{eqn:rvm}
{\rm \psi} = {\rm \psi}_{0} +
{\rm arctan} \left( \frac{{\rm sin}\alpha
\, {\rm sin}(\Phi - \Phi_0 )}{{\rm sin}\zeta
\, {\rm cos}\alpha - {\rm cos}\zeta
\, {\rm sin}\alpha \, {\rm cos}(\Phi - \Phi_0 )} \right),
\end{equation}
where the position angle ($\psi$) increases {clockwise} on the sky.  This definition of $\psi$ is opposite to the astronomical convention (also known as the `observers' convention or the PSR/IEEE convention defined in \citealt{psrchive_ieee}) that $\psi$ increases counterclockwise on the sky, from north to east (cf.~\citealt{DamourTaylor1992, EverettAndWeisberg2001}). A definition of these angles is provided in Fig. \ref{fig:geometry}.

Modelling the position angle swing  of \psr\, using the RVM \citep{Radhakrishnan&Cooke} leads to a solution where the fiducial
plane, $\Phi_0=0$, is located at the central swing underneath the main component. 
The magnetic inclination
angle $\alpha$ and the viewing angle, $\zeta$, are then highly correlated, leading to a solution of small
$\alpha$ and $\zeta$ values (see e.g.~\citealt{Lorimer&Kramer2005}).
Such a solution is inconsistent both with the detection of a Shapiro delay assuming that
the pulsar is spin aligned with the orbital angular momentum vector (i.e.~$\zeta = 180 - i$, \citealt{ksv+21}),
and with the lack of emission over a very wide longitude range, as already discussed in Sect. \ref{sec:xdot}.

Using the information of the Shapiro delay
measurement, one can choose to sample $\zeta$ only from a prior of, say, 88 to 92 deg. Doing so, the resulting
fit still places the fiducial plane underneath the main pulse, but now suggests that the post-cursor's position angles are
separated from the main RVM model by an unusual amount of $\sim$45 deg. Interestingly, recently \cite{Dyks2019} pointed
out that when radio pulsar polarisation is modelled as a coherent sum of natural propagation modes,
for equal amplitudes of these natural propagation modes, two pairs of orthogonal polarisation modes, 
displaced by 45 deg, can be observed. Speculating that the post-cursor emission could result from such conditions,  we modify our fit to include another parameter, an offset in position angle ($\Delta\Psi$) just for the post-cursor, and attempted another blind fit, drawing $\alpha$ and
$\zeta$ from the whole parameter space of 0 to 180 deg. The result is shown in Fig. ~\ref{fig:profile}. Interestingly,
the best fit is now extremely well constrained, and the obtained angles are $\alpha = 95.2 \pm 1.3$ deg and
$\zeta = 88.3 \pm1.3$ deg.  This solution is in excellent agreement with the Shapiro delay measurements, allowing us 
to break the degeneracy and to determine the orbital inclination angle to be $ i = 180 - \zeta = 180 - (88.3 \pm1.3)
= 91.7 \pm 1.3 > 90 $ deg. The value of $\Delta\Psi$ = $48 \pm 3$ deg is consistent with prediction by \cite{Dyks2019} of 45 degrees.  We also point out that this RVM fit would place the second pole at a pulse longitude of
$\sim30$ deg, well separated from the post-cursor component and justifying our notion that it is not an inter-pulse.

Obviously, we are making three major assumptions here. Firstly, we assume that the position angle swing of recycled pulsars has a geometrical origin only and is described by the RVM 
(see \citealt{ksv+21} for a detailed discussion of this assumption).
Secondly, we assume that the spin axis of the pulsar is aligned with the orbital angular momentum, which may not be the case (see Sect. \ref{sec:xdot}).
Finally, we assume that position angle shifts of $\sim$45 deg 
are possible. Turning the argument around, the fact that a blind RVM fit delivers a $\zeta$ value that is in excellent
agreement with the independent timing result, may suggest that the pulsar spin is indeed aligned, Dyks' idea is true and that we should look out for corresponding examples in other pulsars.

A possible alternative solution to the 45 deg problem is that the observed emission does not originate close to the NS in the vicinity of magnetic poles, but is in fact emitted close to the light cylinder with the observed profile strongly influenced by caustic reinforcement. Such an interpretation of the radio emission from high $\dot E$ pulsars, where $E$ is the spin-down luminosity, was proposed by \citet{man05} and \citet{rmh10}, largely motivated by the close relationship of the radio and $\gamma$-ray emission in such pulsars.

\section{Discussion}
\label{sec:discussion}
\begin{figure*}
    \centering
    \includegraphics[angle=270,width=\textwidth]{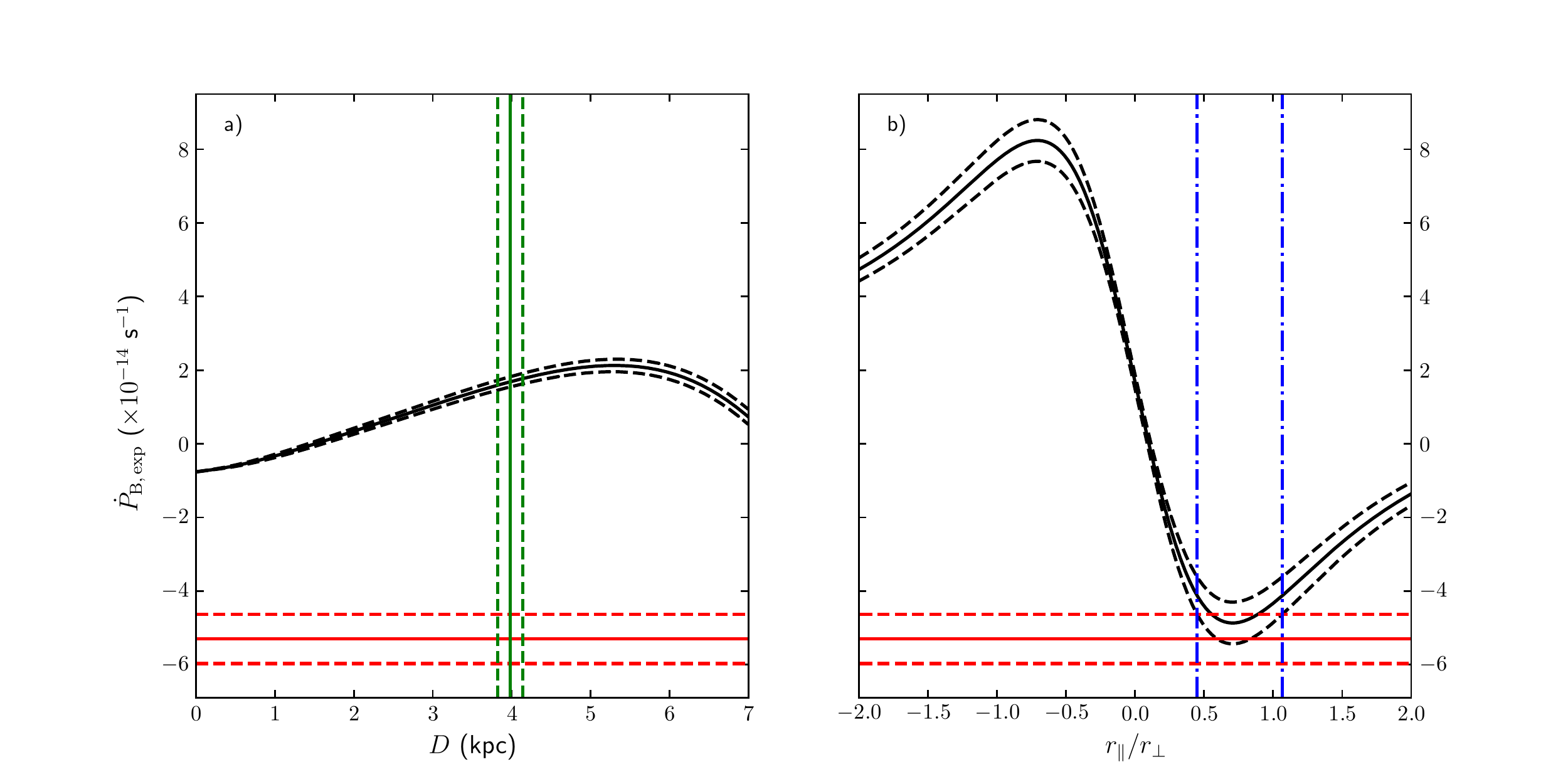}
    \caption{Expected values, $\dot{P}_{\rm B,exp}$ (black), for the
    directly measured time derivative of the orbital period
    in the two scenarios where the \psr\, binary is a field object (panel a)
    or is associated with NGC6752 (panel b). The horizontal scales in panel (a)
    and (b) respectively are the pulsar
    distance, $D$, in kpc, and its depth inside the cluster, $r_\parallel$,
    in units of $r_\perp$. The red line indicates the value $\dot{P}_{\rm
    B,obs}$ measured from our timing. In panel (a) the vertical green line
    indicates the distance of NGC6752 from the Sun. 
    In panel (b) the vertical dot-dashed blue lines delimit the range for
    $r_\parallel$ where $\dot{P}_{\rm B,exp}$ is consistent to $\dot{P}_{\rm B,obs}$. In both panels, dashed lines delimit the 1$\sigma$ 
    limits for the plotted quantities.}
    \label{fig:exppbd}
\end{figure*}

\subsection{The association of \psr\, with the globular cluster NGC6752}
\label{subsec:PSRAassoc}

The  6.37 arcmin offset of \psr\, from the centre of NGC6752 (\citealt{dpf+02}, \citealt{cpl+06}), which corresponds to 7.4\,pc at the cluster distance,  has led some authors to call into question the membership of the pulsar in the GC. As mentioned above, \cite{bkkv06} determined the companion radius from theoretical mass--radius relations for HeWD stars, and derived a distance $D=3.1\pm0.7$\,kpc, which resulted in disagreement with the cluster distance, thus favouring the non-association of \psr\, with NGC6752.

Thanks to our precise measurements of the Shapiro delay (and hence of the masses of the two bodies),  of the proper motion ($\mu = 5.087\pm0.026$\,mas\,yr$^{-1}$) and orbital period derivative ($\dot{P}_{\rm B,obs}=(-5.30_{-0.60}^{+0.74})\times10^{-14}$\,s\,s$^{-1}$) we can now revisit this issue. \cite{bb96} first pointed out that the distance of a binary pulsar located in the Galactic field can be determined by using Eq.\,3.2 from \cite{phi93}, which can be rewritten\footnote{All the quantities in the equation are refereed to measurements performed with respect to the Solar System barycentre.} as

\begin{equation}
\left(\frac{\dot{P}_{\rm B}}{P_{\rm B}}\right)_{\rm obs} =
\left(\frac{\dot{P}_{\rm B}}{P_{\rm B}}\right)_{\rm intr} +
\frac{a_{\rm SHK}}{c} + \frac{a_{\rm MW}}{c},
\label{eq:phifield}
\end{equation}

\noindent
where  $\dot{P}_{\rm B,intr}$  was derived from the component masses in Sect. \ref{subsec:truepdot}, $a_{\rm SHK}\equiv\mu^2D$ is the apparent acceleration of \psr\, due to its transverse motion with respect to the observer, the `Shklovskii' effect \citep{shk70}, and $a_{\rm MW}$ is the component along the line of sight of the true acceleration (hereafter referred to as `acceleration' for simplicity, unless explicitly redefined) imparted on the pulsar by the Milky Way\footnote{Strictly speaking, it is the component along the line of sight of the difference between the accelerations the Milky Way imparts on the pulsar and on the Solar system.}. This acceleration is usually inferred from a model of the Galactic potential and depends on the Galactic coordinates of the pulsar;   both acceleration terms depend on the distance  to the pulsar, $D$. Therefore, if one measures $\mu$ and $\dot{P}_{\rm B}$, and $\dot{P}_{\rm B,intr}$ is independently known, the only unknown quantity in Eq.\,\ref{eq:phifield} is $D$.

We calculated the Milky Way acceleration imparted to \psr\,
by applying Eq.\,16 in \cite{lwj+09}, also using for the vertical component of the Galactic acceleration $F_z$ the analytic formula provided by \citet[their Eq.\,14]{lw21}. The adopted values for the Solar motion in the Galaxy are $\Theta_\odot =240.5\pm4.1$\,km\,s$^{-1}$ and $R_\odot = 8.275\pm0.034$\,kpc \citep{gravcol21}. As a sanity check, we also replaced this model with the ones provided by the python package {\tt GalDynPsr} \citep{pb18} for objects residing at the same Galactic coordinates of \psr, and we found full consistency, in the considered distance range, with the results described below.

Panel (a) of Fig.\,\ref{fig:exppbd} shows the values for the expression in the right hand side of Eq.\,\ref{eq:phifield}, $\dot{P}_{\rm B,exp}$, and along with $\dot{P}_{\rm B,obs}$ for immediate comparison. We considered heliocentric distances up to 7\,kpc, the $5\sigma$ upper limit for the distance of \psr\, derived by \cite{bkkv06}.
It appears that, in the considered distance range, $\dot{P}_{\rm B,exp}$ is never consistent with $\dot{P}_{\rm B,obs}$. This means that the binary system hosting \psr\, must be subjected to a further acceleration component.

The obvious candidate able to generate the extra acceleration is the GC NGC6752. In this case the distance is equal to NGC6752 one, but $\dot{P}_{\rm B,obs}$ must satisfy an expanded version of Eq.\,3.2 from \cite{phi93}:

\begin{equation}
\left(\frac{\dot{P}_{\rm B}}{P_{\rm B}}\right)_{\rm obs} =
\left(\frac{\dot{P}_{\rm B}}{P_{\rm B}}\right)_{\rm intr} +
    \frac{a_{\rm MW}}{c} + \frac{a_{\rm SHK}}{c} +
    \frac{a_{\rm GC}}{c}
    \label{eq:phi93}
,\end{equation}

\noindent and the acceleration $a_{\rm GC}$ imparted by the cluster can be expressed as
\begin{equation}
  a_{\rm GC}=-GM(r_{\rm PSRA})
  \frac{r_\parallel}{(r_{\perp}^2+r_\parallel^{2})^{3/2}},
\label{eq:accel}
\end{equation}

\noindent
where $G$ is Newton's gravitational constant, $M(r_{\rm PSRA})$ is the
cluster mass enclosed in a radius equal to $r_{\rm PSRA}$ (i.e. the \psr\, distance from the cluster centre), and $r_{\perp}$ and $r_\parallel$ are the pulsar coordinates in the cluster frame, perpendicular and parallel to the line of sight, respectively.

Given the high displacement of \psr\, with respect to the cluster centre, we assumed negligible the cluster mass outside $r_{\rm PSRA}$, hence $M(r_{\rm PSRA})=M_{\rm NGC6752}=(2.76\pm0.04)\times10^5M_\odot$ \citep{hbsb20}. At the cluster distance $D=3.984\pm0.159$\, kpc, the accelerations due to the Milky Way and the \citeauthor{shk70} effect are $a_{\rm MW}=(+2.64\pm0.74)\times10^{-11}$\,m\,s$^{-2}$
($a_{\rm MW}/c=(+8.8\pm2.5)\times10^{-20}$\,s$^{-1}$)
and
$a_{\rm SHK}=(7.5\pm0.3)\times10^{-11}$\,m\,s$^{-2}$
($a_{\rm SHK}/c=(2.5\pm0.1)\times10^{-19}$\,s$^{-1}$),
respectively. 
Panel (b) of Fig.\,\ref{fig:exppbd} displays $\dot{P}_{\rm B,exp}$ after taking into account the acceleration imparted by the cluster, and shows that $\dot{P}_{\rm B,exp}$ is consistent with our measured value if the \psr\, depth in the cluster lies in the range $0.45\leq r_\parallel /r_\perp \leq 1.07$ (1$\sigma$  level), thus implying that NGC6752 can well be the object responsible for the observed needed extra acceleration on the \psr\, binary.

\begin{figure}
    \centering
    \includegraphics[angle=270,width=8cm]{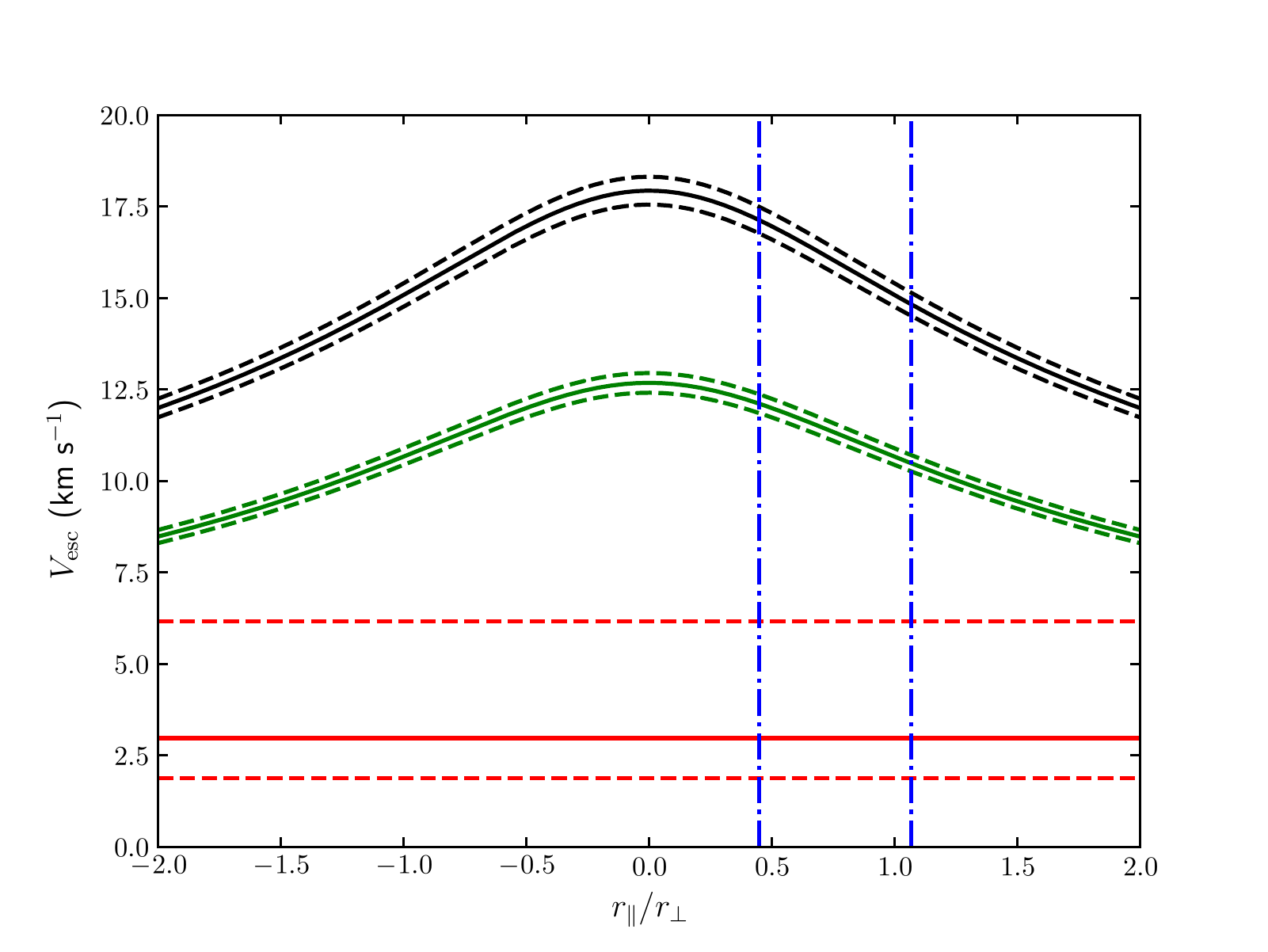}
    \caption{NGC6752 escape (black) and circular orbit (green) velocity at
    the \psr\, projected position, $r_\perp$, versus its depth inside the cluster, $r_\parallel$, plotted in units of $r_\perp$. The red line indicates the pulsar
    3D velocity in the cluster's frame. Dashed lines delimit the 1$\sigma$
    limits for the plotted quantities. The vertical dot-dashed blue lines 
    mark the range for $r_\parallel$ where the excess of acceleration can be explained as due to NGC6752 (see Fig.\ref{fig:exppbd}, panel b).}
\label{fig:escape}
\end{figure}

We also compared the \psr\, 3D velocity in the frame of NGC6752 with the cluster escape velocity, $V_{\rm esc}(r_{\rm PSRA}),$ at the pulsar position:
\begin{equation}
  V_{\rm esc}(r_{\rm PSRA}) = \sqrt{\frac{2GM(r_{\rm PSRA})} {r_{\rm PSRA}}},
\label{eq:escape}
\end{equation}
\noindent
where all symbols in the right hand side of Eq.\,\ref{eq:escape} are defined as in Eq.\,\ref{eq:accel}. \psr's motion in the cluster frame, after considering the measured radial velocity and proper motion of the cluster from the \textit{Gaia} Early Data Release 3 data \citep{vas19,vb21}, and the binary radial velocity $V_{\rm R}=-28.1\pm4.9$\,km\,s$^{-1}$ \citep{cfpd06}, is given by $\Delta\mu_{\alpha}\cos\delta = -0.069\pm0.030$\,mas\,yr$^{-1}$, $\Delta\mu_\delta = +0.097\pm0.037$\,mas\,yr$^{-1},$ and $\Delta V_{\rm R} = -1.82\pm4.9$\,km\,s$^{-1}$.  At the cluster distance of $3.98\pm0.16$\,kpc this implies a relative 3D velocity 
$V_{\rm 3D}=3.0_{-1.1}^{+3.2}$\,km\,s$^{-1}$.
We compare in Fig.\,\ref{fig:escape} the \psr\, 3D velocity $V_{\rm 3D}$ in the cluster frame to $V_{\rm esc,PSRA}$ as a function of the pulsar depth $r_\parallel$ in the GC. In the range for $r_\parallel$ where the observed acceleration on \psr\, can be ascribed to NGC6752 (see Fig.\,\ref{fig:exppbd}), $V_{\rm 3D}$ is never larger than the cluster escape velocity, thus  demonstrating that \psr\, is bound to the cluster. Moreover, Fig.\,\ref{fig:escape} shows that \psr's 3D velocity in the cluster's frame is also lower than the circular orbital velocity at the  pulsar position: this means that \psr\, is also falling (back) towards the centre of NGC6752. Therefore, our analysis of the measured time derivative of the \psr\, orbital period, coupled with the determination of the masses of the two stars in the binary, allows us to build up a consistent scenario where this binary is located in the outskirts of the GC NGC6752, and gravitationally bound to it. Thus, we can finally unambiguously conclude that the pulsar \psr\, is associated with the GC NGC6752.

\subsection{Testing WD models with \psr}
\label{subsec:wdmodels}

\psr's companion is a WD for which a large set of parameters are known. Their list and values are presented in Table\,\ref{tab:wdpars}. In analogy to the tests on gravity theories performed with pulsars in relativistic binaries, where systems for which at least three post--Keplerian parameters allow a test of the validity of the assumed gravity theory (see e.g. \citealt{ksm+21}), we can use the set of \psr's known parameters to test models for the structure and evolution of WDs.

\begin{table}[h]
    \caption{Model-independent measured parameters for the HeWD binary companion of \psr.}
   \centering
    \begin{tabular}{llc|l}
    \hline
Parameter & Symbol & Value & Ref. \\
\hline
Mass ($M_\odot$) & $M_{\rm C}$ & $0.202\pm0.006$ & a\\
Distance (kpc) & $D$ & $3.984\pm0.159$ & a$^{\it 1,2}$\\
Reddening (mag) & $E(B-V)$ & $0.046\pm0.005$ & a$^{\it 1,3}$\\
U--band magnitude (mag) &$m_{\rm U}$&$22.02\pm0.05$& b\\
B--band magnitude (mag) &$m_{\rm B}$&$22.22\pm0.03$& b\\
V--band magnitude (mag) &$m_{\rm V}$&$22.13\pm0.02$& b\\
Surface temperature (K)  & $T_{eff}$ & $10090\pm150$& b\\
Surface gravity (c.g.s.)& $\log_{10} g$& $6.44\pm0.22$&b\\
\hline
    \end{tabular}
    \label{tab:wdpars}
    \begin{tablenotes}
    References: a: this work; b: \cite{bkkv06}\\
      ({\it 1}) Consequence of our demonstration that
      \psr\\ is member of NGC6752.\\
      ({\it 2}) NGC6752 distance \citep{vb21}.\\
      ({\it 3}) NGC6752 reddening \citep{gbc+05}.\\
    \end{tablenotes}
\end{table}

\begin{figure}[ht]
    \centering
    \includegraphics[angle=0,width=0.50\textwidth]{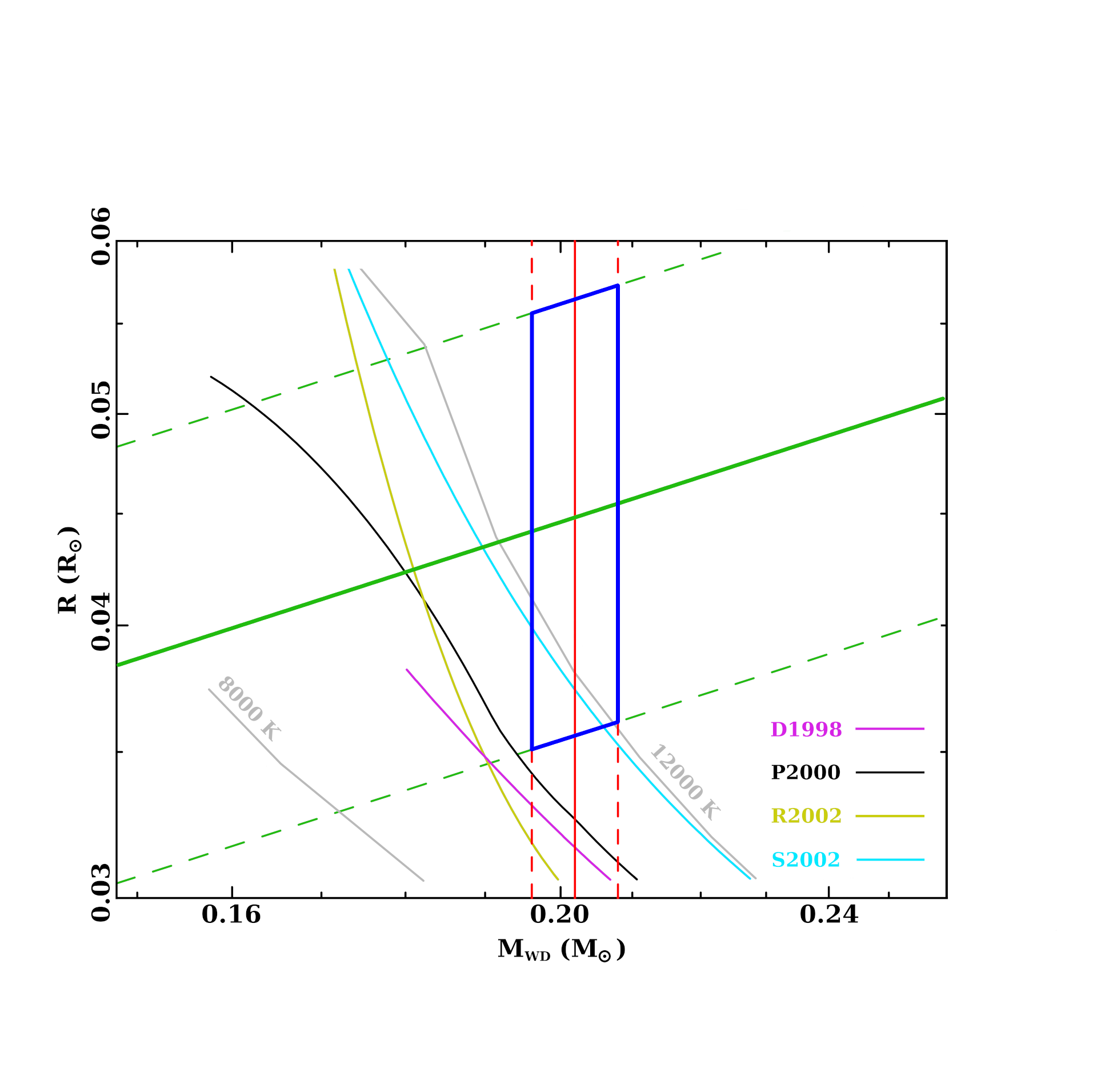}
    \caption{M--R relations derived by \citet{bkkv06}, and already presented in their Fig.\,6, obtained from the following theoretical works: \citeauthor{dsbh98} (\citeyear{dsbh98}; magenta; D1998); \citeauthor{pab00} (\citeyear{pab00}; black: $T_{\rm eff} = 10090$\,K, light grey: $T_{eff} = 8000$ and $12000$\,K; P2000); {\citeauthor{rsab02} (\citeyear{rsab02}; dark yellow; R2002)}, and \citeauthor{sarb02} (\citeyear{sarb02}; cyan; S2002). The diagonal green and vertical red lines mark the surface gravity measured by \cite{bkkv06} and our measurement of the PSR\,J1910-5959A companion's mass, respectively (solid: best values, dashed: 1$\sigma$ lower and upper limits). The blue parallelogram delimits the region in the M--R space consistent with the measured surface gravity and companion mass.}
\label{fig:mrb06}
\end{figure}

A first kind of test can be done on the theoretical mass-radius (M--R) relations of WDs. For instance, we can consider Fig.\,\ref{fig:mrb06}, a reproduction of Fig.\,6 from \cite{bkkv06}, which shows M--R curves that result from some theoretical models of HeWD with $T_{\rm eff}=10090\pm150$\,K, the measured effective temperature of \psr\, companion.
If we put our measurement of the companion mass into that context, it results that the M--R relation by \cite{sarb02} is the only one consistent (at 1$\sigma$) with the surface gravity measured by \cite{bkkv06}. The \citeauthor{sarb02} model implies a radius
$R_{\rm C}=0.0376^{+0.0024}_{-0.0013}\,R_{\odot}$,
which is inconsistent with the value $R_{\rm C}=0.058\pm0.004\,R_{\odot}=(4.03\pm0.28)\times10^4$\,km \citeauthor{bkkv06} obtained from the measured optical flux and the intrinsic one predicted by WD models, under the assumption \psr\, is at the same distance of NGC6752.  It is worth mentioning that from our mass measurement the model by \cite{pab00} for $T_{\rm eff}=12000$\,K is also valid and returns a WD radius fully consistent with the value inferred by the \citeauthor{sarb02} M$-$R relation. Nevertheless, by using the surface gravity and temperature obtained by \cite{bkkv06}, \cite{amc13} calculated $M_{\rm C}=0.185\pm0.0041M_{\odot}$ and a cooling age $\tau_{\rm c} = 1.47\pm0.17$\,Gyr. The value of the mass is consistent only at the 2.1$\sigma$ level with our determination.

\begin{figure}[ht]
    \centering
    \includegraphics[angle=270,width=0.49\textwidth]{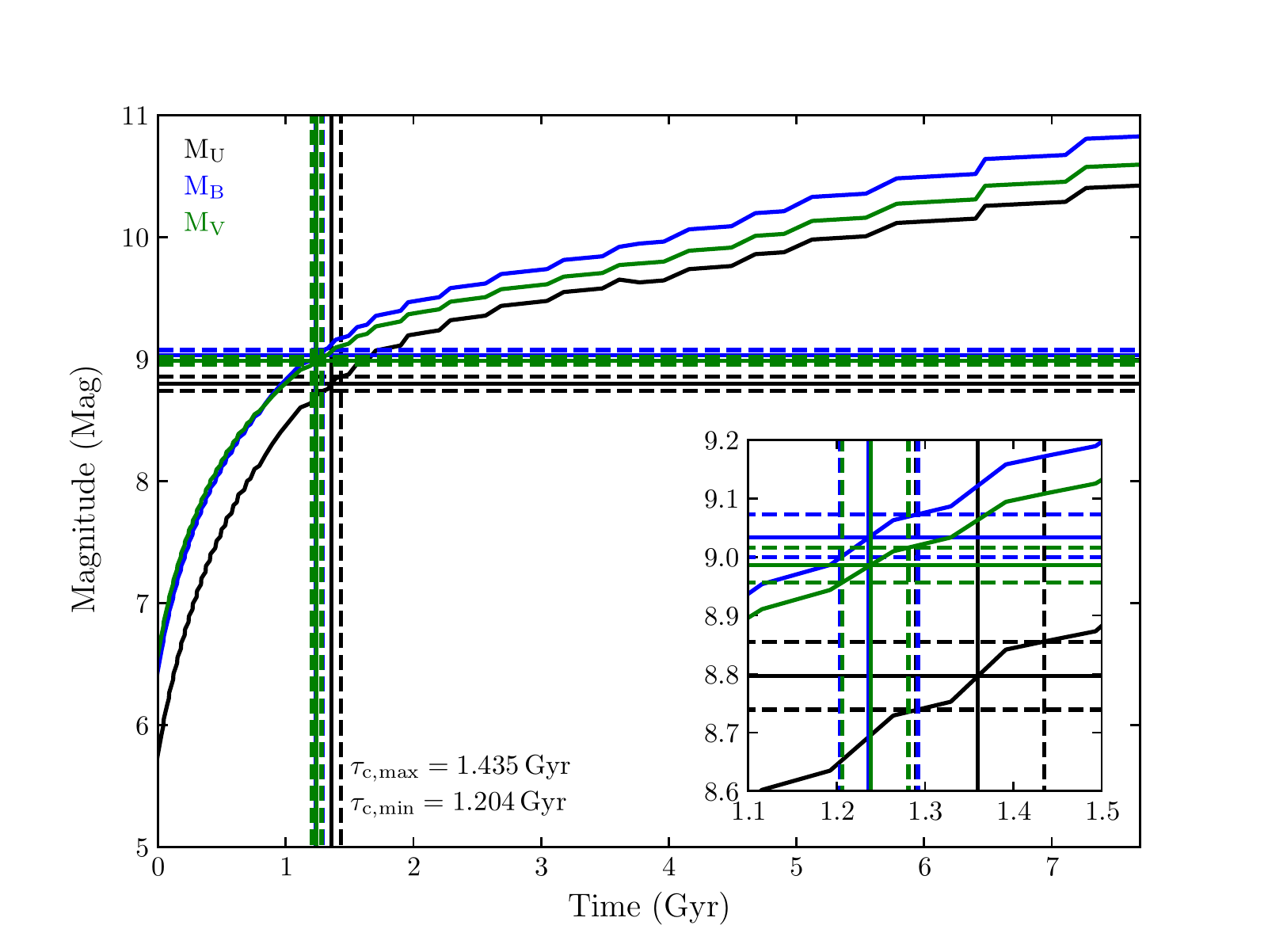}
    \caption{UBV absolute magnitudes (black, blue, and green lines) versus time as predicted by the \cite{imt+16} evolution track. Horizontal lines mark the magnitude of the \psr\, companion in each band, and vertical lines mark
    the corresponding inferred cooling age. Dashed lines delimit 1$\sigma$ uncertainty ranges. The inset displays a zoomed-in view of the time and magnitude ranges where the cooling age of the WD companion is inferred from its measured apparent magnitudes converted to absolute magnitudes.}
\label{fig:wdmag1}
\end{figure}

\begin{figure}[ht]
    \centering
    \includegraphics[angle=270,width=0.49\textwidth]{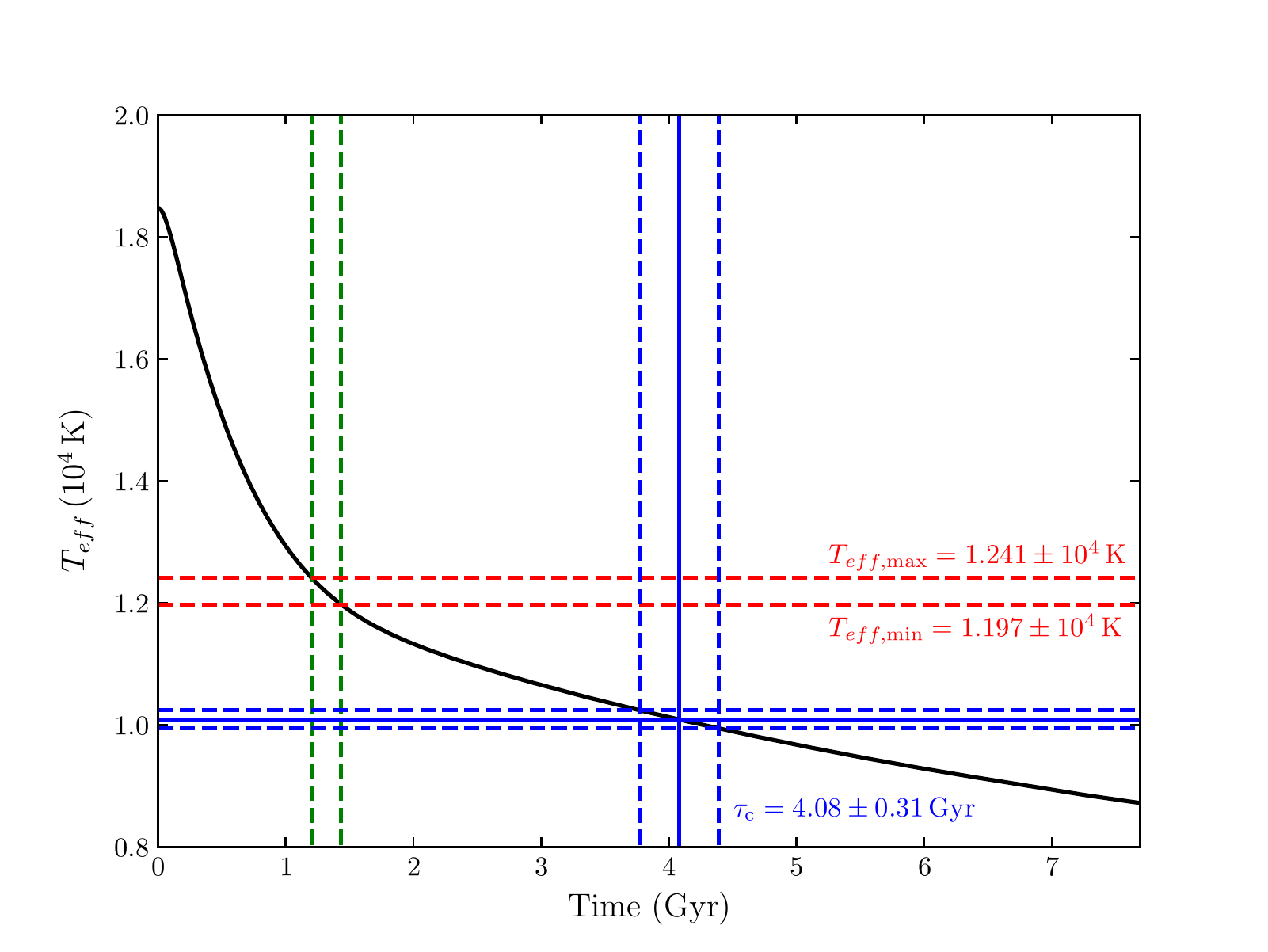}
    \caption{WD effective temperature versus time as predicted by the \cite{imt+16} evolution track. The vertical dashed green lines delimit the age range as inferred from Fig.\,\ref{fig:wdmag1}, and the red horizontal ones the corresponding WD $T_{eff}$ range. The horizontal blue line indicates the effective temperature measured by
    \cite{bkkv06}, while the vertical one the corresponding cooling age. Dashed blue lines delimit 1$\sigma$ uncertainty ranges.}
\label{fig:wdtemp}
\end{figure}

\begin{figure}[ht]
    \centering
    \includegraphics[angle=270,width=0.49\textwidth]{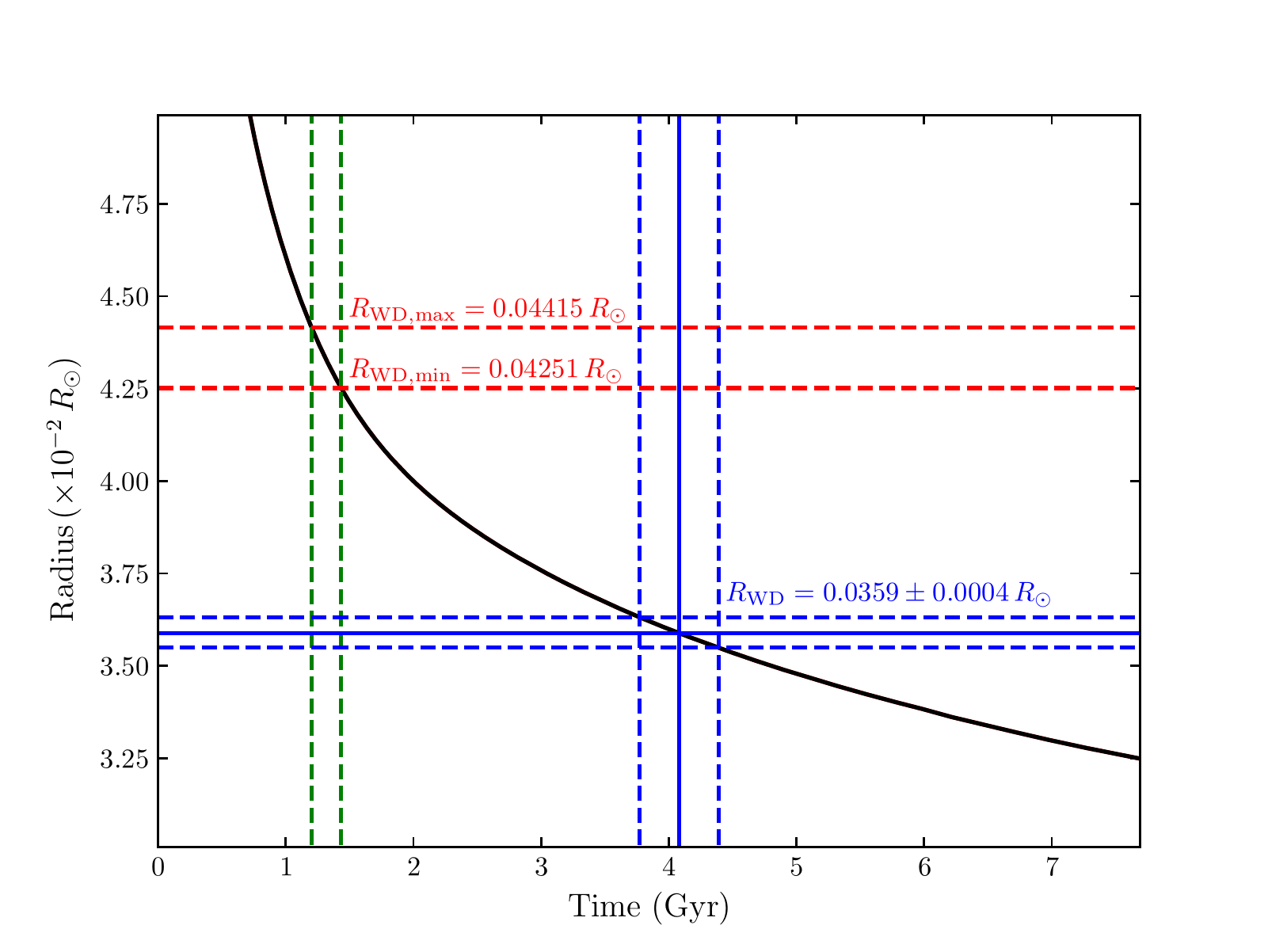}
    \caption{WD radius versus time as predicted by the \cite{imt+16} evolution track. The vertical dashed green lines delimit the age range as inferred from Fig.\,\ref{fig:wdmag1} and the red horizontal ones the corresponding WD radius range. The vertical blue line indicates the cooling age deduced from Fig.\,\ref{fig:wdtemp}, while the blue horizontal one marks the corresponding WD radius. Dashed blue lines delimit 1$\sigma$ uncertainty ranges.}
\label{fig:wdradius}
\end{figure}

A second kind of test can be performed on theoretical time evolution tracks of WD parameters. Recently, evolution tracks for HeWDs in NS$-$WD systems have been computed \citep{imt+16,cjp+20} for a large set of values of the HeWD mass and metallicity, under the form of tabulated values of several WD parameters as a function of time. Such models can be tested by deriving the WD age from the evolution of one selected parameter, and checking whether the model, at the obtained age, predicts for other parameters a value in agreement with the measured one. Given our measurement of the companion mass and the metallicity of NGC6752 ([Fe/H]$=-1.43\pm0.04$, \citealt{gbc+03}), we picked the track for a HeWD with a mass $M_{\rm WD}=0.202M_\odot$ and a metallicity $Z_{\rm  WD}=0.0005$ against the known features of the \psr\, companion. We first selected the luminosity evolution (i.e. the WD magnitude) as the age indicator, and obtained a cooling age $\tau_{\rm c}$ in the range $1.2{\rm Gyr}\leq\tau_{\rm c}\leq1.44{\rm Gyr}$ (see Fig.\,\ref{fig:wdmag1}), after converting the U, B, and V apparent magnitudes reported in \cite{bkkv06} to absolute magnitudes. The inferred value for $\tau_{\rm c}$ is consistent with the one obtained by \cite{amc13}, but the corresponding effective temperature $1.20\times10^4\leq T_{\rm eff} / K \leq 1.24\times10^4$ (see Fig.\,\ref{fig:wdtemp}) is not consistent with the value measured by \cite{bkkv06} via spectroscopy.
Instead, the corresponding WD radius, $0.0425\leq R_{\rm WD}/R_\odot \leq 0.0442$ (see Fig.\,\ref{fig:wdradius}), is in agreement with the radius value derived by \cite{bkkv06} from the M--R tracks they considered.
We also considered the effective temperature as age indicator: from the value $T_{\rm eff}=10090\pm150$\,K, \citep{bkkv06} we deduce $\tau_{\rm c}=4.08\pm0.31$\,Gyr (see Fig.\,\ref{fig:wdtemp}), in disagreement with the age inferred from the luminosity evolution.
Nevertheless, the predicted WD radius at this age is $R_{\rm C}=0.0359\pm0.0004$ (see Fig.\,\ref{fig:wdradius}), which is consistent with the value we obtained from the M--R relation by \cite{sarb02}.

In summary, all currently considered theoretical works about the structure and the evolution of a HeWD in a MSP$-$WD binary are based on hypothesis that are capable to capture some of the measured parameters of NGC6752A companion, but none of them consistently predicts them all under a single framework.

\subsection{Alignment of the WD rotation}

 As shown in Sect.~\ref{sec:xdot} and calculated in detail in the appendix, the only way to explain $\dot{x}_{\rm obs}$ is via spin-orbit coupling, caused mostly by the quadrupole moment of the He WD companion. This requires a misalignment between the spin of the WD and the angular momentum of the orbit. As mentioned earlier, this is unexpected: during the evolution of a low-mass X-ray binary (LMXB), the mass transfer that spins up the pulsar and circularises the system also aligns the spin of the component stars with the orbital angular momentum.

This expectation assumes that the LMXB evolution was not perturbed by close encounters with other stars. This is not the case in GCs: the known population of binary pulsars in GCs shows plenty of evidence for close encounters, either from the abnormally large orbital eccentricities of MSP--HeWD systems or from eccentric MSP binaries with massive compact companions, which result from exchange encounters involving MSPs (e.g. \citealt{1991ApJ...374L..41P,2004ApJ...606L..53F,2012ApJ...745..109L}). Such close encounters can change the orbital plane of a binary, but if that happens, then it is almost inevitable that they increase orbital eccentricity of the system by orders of magnitude \citep{1992RSPTA.341...39P}. The low orbital eccentricity of the \psr\, system ($e \sim 8 \times 10^{-7}$), which is typical of MSP--HeWD systems in the Galactic disk with this orbital period \citep{1992RSPTA.341...39P}, seems to make this scenario unlikely.

However, the \psr\, system is found far from the centre of NGC~6752 but bound to it, and it likely has a nearly radial orbit around the centre of the GC, as we established in Sect.~\ref{subsec:PSRAassoc}. This means two things: that it was likely near the core like most other pulsars in this cluster, and that it was later almost ejected from the cluster after a very close encounter. This implies that almost certainly there was a significant perturbation in the history of this system, even though the orbital eccentricity does not reflect this.

This leads us to the question of how we reconcile the evidence of violent events in the past -- the position of the system relative to GC, and now the apparent misalignment in the WD rotation -- with the low eccentricity. A possibility, proposed by \cite{2002ApJ...570L..85C}, was that the system was ejected by interaction with a massive binary black hole.
Such an interaction could in principle produce a large change of velocity of the system but, because of the low tides involved, it would not significantly change its orbital eccentricity. However, such an interaction would also not change the orbital orientation of the system, which would mean that in this case there should be no misalignment of the WD spin. 

We propose instead that (a) the close encounter that nearly ejected the system was a more normal type of encounter with a much less massive star, which would have changed the orbital eccentricity and the orbital plane, producing the current WD misalignment, and (b) the system was later circularised without aligning the WD rotation. This could happen, for instance, if the encounter happened soon after the LMXB phase, when the WD was hot enough to be bloated to very large radii by Hydrogen shell flashes. This bloated atmosphere circularised the orbit.

However, it is currently not clear to us whether such a scenario - circularising the orbit without aligning the WD rotation - is even possible; this would require a detailed binary evolution / perturbation simulation that is clearly beyond the scope of this paper. Investigating such scenarios would almost lead to an improved understanding of the evolutionary history of this intriguing system.

Future optical observations might also further constrain the radius and potentially the spin period of the WD. Such observations would be able to test this hypothesis and to finally assess the nature of the main contribution to the $\dot{x}_{\rm obs}$ detected in \psr.

\section{Summary}
\label{sec:summary}

We have reported on the analysis of $\sim22$\,years of observations of the binary pulsar \psr\, in the GC NGC6752, conducted with the Parkes 64 m Murriyang and MeerKAT radio telescopes. The full Stokes observations with MeerKAT allowed us to investigate the shape and polarimetry of the pulsar profile and thus obtain the rotation measure along the line of sight up to \psr. However, we did not find any evidence of signal scattering in the ionised ISM. Thanks to the large time span covered by the observations with the Parkes radio telescope and the outstanding sensitivity of the MeerKAT radio telescope, we have measured several orbital and post-Keplerian parameters with a greatly improved precision. 

We used the measurement of the Shapiro delay to infer precise masses for the pulsar and companion, and found them to be very consistent with the deduction of masses from optical observations of the WD companion.  We measured a secular decay of the orbital period and used it to not only derive the true spin period derivative of the pulsar, but also to prove that the measured value can only be explained by invoking apparent changes in the orbital period caused by the acceleration in the GC. This confirms, incidentally, that the pulsar indeed belongs to NGC6752, thereby settling a long-standing debate about its association. We interpreted our $5\sigma$ measurement of the rate of change of the pulsar's projected semi-major axis in terms of spin$-$orbit coupling of the WD companion.
This requires the rotation of the companion WD to be misaligned with the orbital angular momentum. This could have been caused by a violent interaction of this system with another star -- possibly the one that almost ejected this binary out of NGC~6752; however, that scenario is difficult to reconcile with the low observed orbital eccentricity. We discussed several possible solutions to this problem. Future optical observations that constrain the spin period of the WD might allow the idea that the WD spin is misaligned with the orbital angular momentum to be tested.

Our analysis of the pulsar's polarisation using the RVM and just the position angle of the main pulse provided interesting evidence that the post-cursor position angle is shifted by exactly $45\deg$ from the expected position, providing rare evidence for coherent mixing of two equal-amplitude natural propagation modes. This motivates further investigation of this phenomenon in other pulsars and supports our idea that the post-cursor emission does not arise from the opposite pole of the pulsar. 
A possible alternative solution to the 45 deg problem is that the observed radio pulses are generated close to the light cylinder, with their form strongly influenced by caustic reinforcements.

Finally, the very high precision of the WD mass measurement, jointly with other parameters measured for this object with optical observations and reported in literature, allowed us to show how this system can be used as a test bed for structural models of WDs and evolutionary models of WD--NS binaries in GCs.

\begin{acknowledgements}
The authors would like to thank Mario Cadelano, Alina Istrate, Norbert Langer, Thomas Tauris, Norbert Wex, Kent Yagi and Sophia Yi for insights and valuable discussions on the evolution of the binary system and on WD models. We thank Cees Bassa, Marcus Lower and Sarrvesh Sridhar for comments on parts of the manuscript. The MeerKAT telescope is operated by the South African Radio Astronomy Observatory, which is a facility of the National Research Foundation, an agency of the Department of Science and Innovation. SARAO acknowledges the ongoing advice and calibration of GPS systems by the National Metrology Institute of South Africa (NMISA) and the time space reference systems department department of the Paris Observatory. MeerTime data is housed on the OzSTAR supercomputer at Swinburne University of Technology supported by ADACS and the Gravitational Wave Data Centre via Astronomy Australia Ltd. The Parkes radio telescope is funded by the Commonwealth of Australia for operation as a National Facility managed by CSIRO. We acknowledge the Wiradjuri people as the traditional owners of the Observatory site. The National Radio Astronomy Observatory is a facility of the National Science Foundation operated under cooperative agreement by Associated Universities, Inc. This research has made extensive use of NASAs Astrophysics Data System (https://ui.adsabs.harvard.edu/) and includes archived data obtained through the CSIRO Data Access Portal (http://data.csiro.au). Parts of this research were conducted by the Australian Research Council Centre of Excellence for Gravitational Wave Discovery (OzGrav), through project number CE170100004. VVK, PCCF, MK, AP, WC, AR, FA, EDB, VB, DJC and PVP acknowledge continuing valuable support from the Max-Planck Society. APo and AR acknowledge the support from the Ministero degli Affari Esteri e della Cooperazione Internazionale - Direzione Generale per la Promozione del Sistema Paese - Progetto di Grande Rilevanza ZA18GR02. APo and AR acknowledge support through the research grant "iPeska" (PI: Andrea Possenti) funded under the INAF national call Prin-SKA/CTA approved with the Presidential Decree 70/2016. AK acknowledges funding from the UK Science and Technology Facilities Council (STFC) consolidated grant to Oxford Astrophysics, code ST/000488. This publication made use of open source python libraries including Numpy \citep{numpy}, Matplotlib \citep{matplotlib}, Astropy \citep{astropy} and Chain Consumer \citep{Hinton2016_ChainConsumer}, BILBY \citep{bilby} and Dynesty \citep{dynesty} along with pulsar analysis packages: \textsc{psrchive} \citep{HotanEtAl2004}, \textsc{tempo2} \citep{HobbsEtAl2006}, \textsc{temponest} \citep{LentatiEtAl2014}.
\end{acknowledgements}

\bibliographystyle{aa}
\bibliography{journals.bib,main.bib,global.bib}

\clearpage
\newpage
\section*{Appendix: Contributions to $\dot{x}$ from the change in the aberration parameter and spin-orbit coupling}

\subsection*{Change of the aberration parameter}

In this section we discuss how the contribution from $\dot{\epsilon}_{\rm A}$ to $\dot{x}_{\rm obs}$ is likely to be insignificant. The pulsar contribution from $\dot{\epsilon}_{\rm A}$ can be expressed as 
\begin{equation}
\dot{x}_{\dot{\epsilon}_{\rm A}} 
= x\left(\frac{d\epsilon_{\rm A}}{dt}\right) 
= -x\,\frac{P_{\rm 0}}{P_{\rm B}} \, \frac{\Omega_{\rm geod}^{\rm p}}{(1- e^2)^{1/2}}\, 
  \frac{\cot \lambda_{\rm P} \sin 2\eta_{\rm P} + \cot \textit{i} \cos \eta_{\rm P}}{\sin \lambda_{\rm P}},
  \label{eq:dotxe}
\end{equation}
where $\theta_{\rm P}$ is the longitude of precession,  $\lambda_{\rm P} \equiv 180 - \zeta \equiv 180- (\alpha_{\rm P} + \beta_{\rm P})$ is the polar angle of the pulsar spin axis with $\alpha_{\rm P}$ its magnetic inclination and $\beta_{\rm P}$ the impact angle of our line of sight to the pulsar emission cone \citep{DamourTaylor1992, Lorimer&Kramer2005}. $\Omega_{\rm geod}^{\rm P}$ is the rate of geodetic precession of the pulsar, which is given by
\begin{equation}\label{eq:geod}
\Omega_{\rm geod}^{\rm P} = \left(\frac{2\pi}{P_{\rm B}}\right)^{5/3}  {T_\odot}^{2/3} M_{\rm C} \, \frac { (4M_{\rm P} + 3M_{\rm C})}{2M_{\rm TOT}^{4/3}} \frac{1}{1-e^2}
\end{equation}
\citep{Lorimer&Kramer2005}. Our measurements give $\Omega_{\rm geod}^{\rm P} \sim 0.02 \deg~ \rm yr^{-1}$, which implies a total precession of $\sim 0.4 \deg$ over the course of our dataset. All angles, like the measurement of the Keplerian parameters, are with respect to the reference epoch, $T_{0}$.

\begin{figure}
    \centering
    \includegraphics[width = 0.49
    \textwidth]{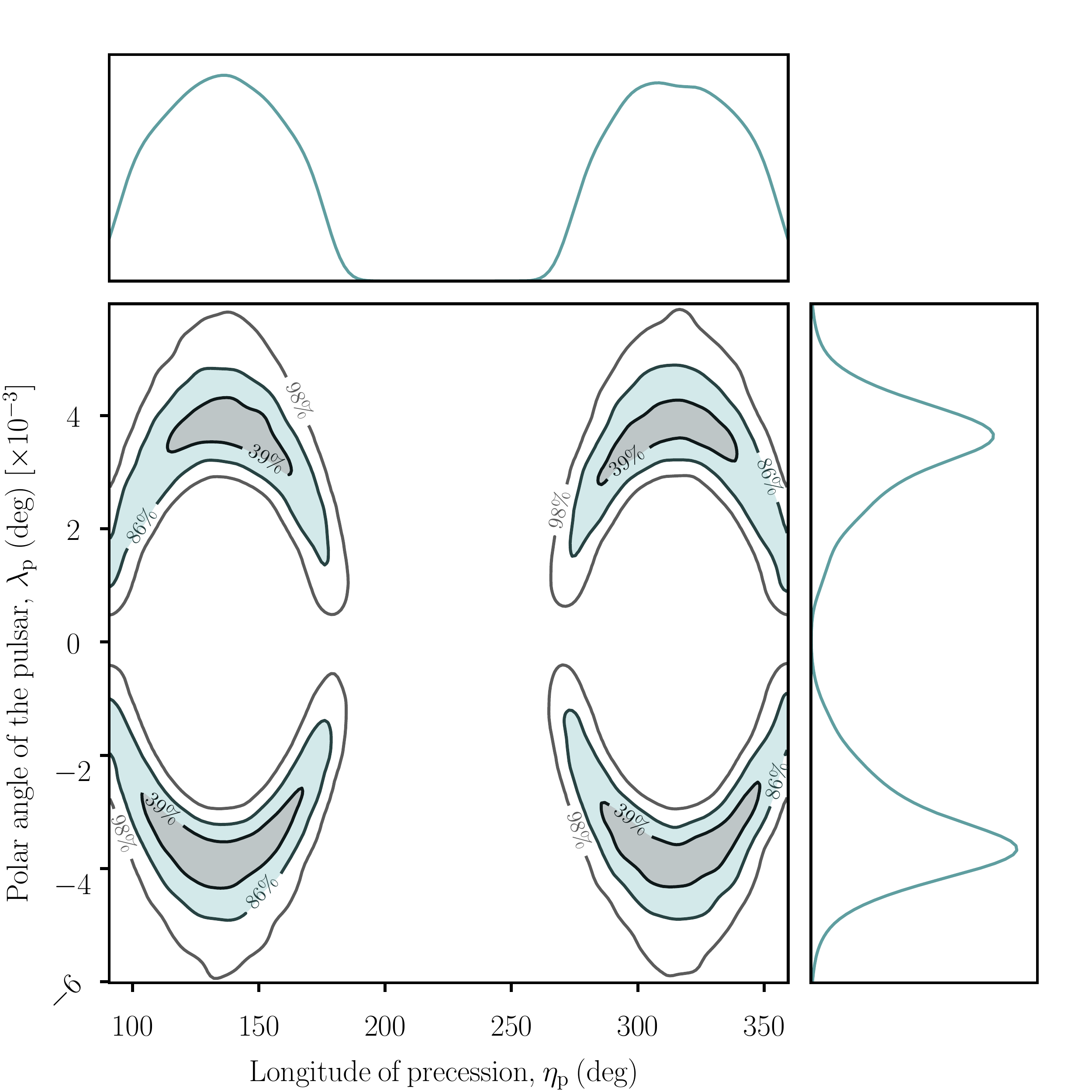}
    \caption{ Constraints on the longitude of precession of the pulsar ($\theta_{\rm P}$) and the polar angle of the pulsar spin axis ($\lambda_{\rm P}$) with respect to our line of sight, if all of $\dot{x}_{\rm obs}$ is contributed by the changing aberration of the pulsar. Note that these constraints are independent of the sense of the inclination angle. See text for more details.
    }
\label{fig:aberration}
\end{figure}

Figure\,\ref{fig:aberration} shows the constraints on (${\lambda_{\rm P}, \theta_{\rm P}}$) that would contribute to the entirety of $\dot{x}_{\rm obs}$. We obtain a very tight constraint of $0.002 \lapp |\lambda_{\rm P}| \lapp 0.006 \deg$ regardless of the sense of the inclination angle. Such a small value of  $\lambda_{\rm P}$ is {a priori} unlikely if one assumes a random orientation of the spin axis about our line of sight, for which the prior probability density function equals $\sin \lambda_{P}$. This makes it unlikely for  $\dot{\epsilon}_{\rm A}$ to significantly contribute to $\dot{x}_{\rm obs}$. 

Furthermore, such a low value of $\lambda_{\rm P}$ means that the magnetic axis is almost perfectly aligned with the spin axis of the pulsar. Such a configuration should ideally produce a pulse profile with a nearly 100\% duty cycle (assuming that the beam is filled), whereas we see that the on-pulse region of the pulse profile nominally covers less than half of the rotational phase. It is important to note that while there are two widely separated, distinct pulses in the pulse profile that we term them as pulse and post-cursor (see Sect. \ref{sec:profile}), they are not separated by $180 \deg$ as one would expect if the emission were from two opposite poles. This raises the suspicion that these pulses are part of a much larger, very patchy, aligned emission cone. This idea is further discussed in \S \ref{sec:rvm} and is considered unlikely. 

Finally, the alignment of the spin and magnetic axis, combined with the Shapiro delay measurement, implies that the misalignment angle of the pulsar's spin  from the orbital angular momentum is almost $90 \deg$. This means that the rate of change of the longitude of precession, $\eta_{\rm P}$ (and hence $\lambda_{\rm P}$) is roughly the same as $\Omega_{\rm geod}^{\rm P}$. Hence, even if it were true that at $t= T_{0}$, $\lambda_{\rm P} \sim 0$, it will precess away to $\sim \pm 0.4 \deg,$  which would then produce a considerably minuscule contribution to $\dot{x}_{\rm obs}$. Consequently, the evolution of $x$ for such a configuration would be more complicated and rapid than the simple formula given by Eq. \ref{eq:dotxe} and would have given rise to higher-order secular contributions to changes in $x,$ which we do not see (our fits yield results consistent with zero). All these arguments point to the fact that changing aberration cannot be the sole or even a dominant contributor to $\dot{x}_{\rm obs}$. 

The very small inferred precession rate hinders finding any observable change in the pulse profile of the pulsar over our timing baseline of 22 years, which would otherwise be clear proof for the misalignment of the pulsar's spin, and could have provided additional constraints on the pulsar geometry. This is complicated by the fact that (due to this being a MSP in a GC hosting several other MSPs), most of the data were taken in search mode with just total intensity, incoherent dedispersion and too coarse a time resolution. Hence, folding the data only provides 64 independent pulsar phase bins implying a bin resolution of $\sim 5.6 \deg$ in pulse longitude, not enough to be sensitive to small profile changes due to geodetic precession. On the other hand, we also compared the MeerKAT observations to some Parkes data taken in 2013 at higher time resolution, thus allowing  the pulse profile to be resolved into 512 bins: the corresponding total intensity profiles appear identical. 

\begin{figure*}
    \centering
    \includegraphics[width = 0.9\textwidth]{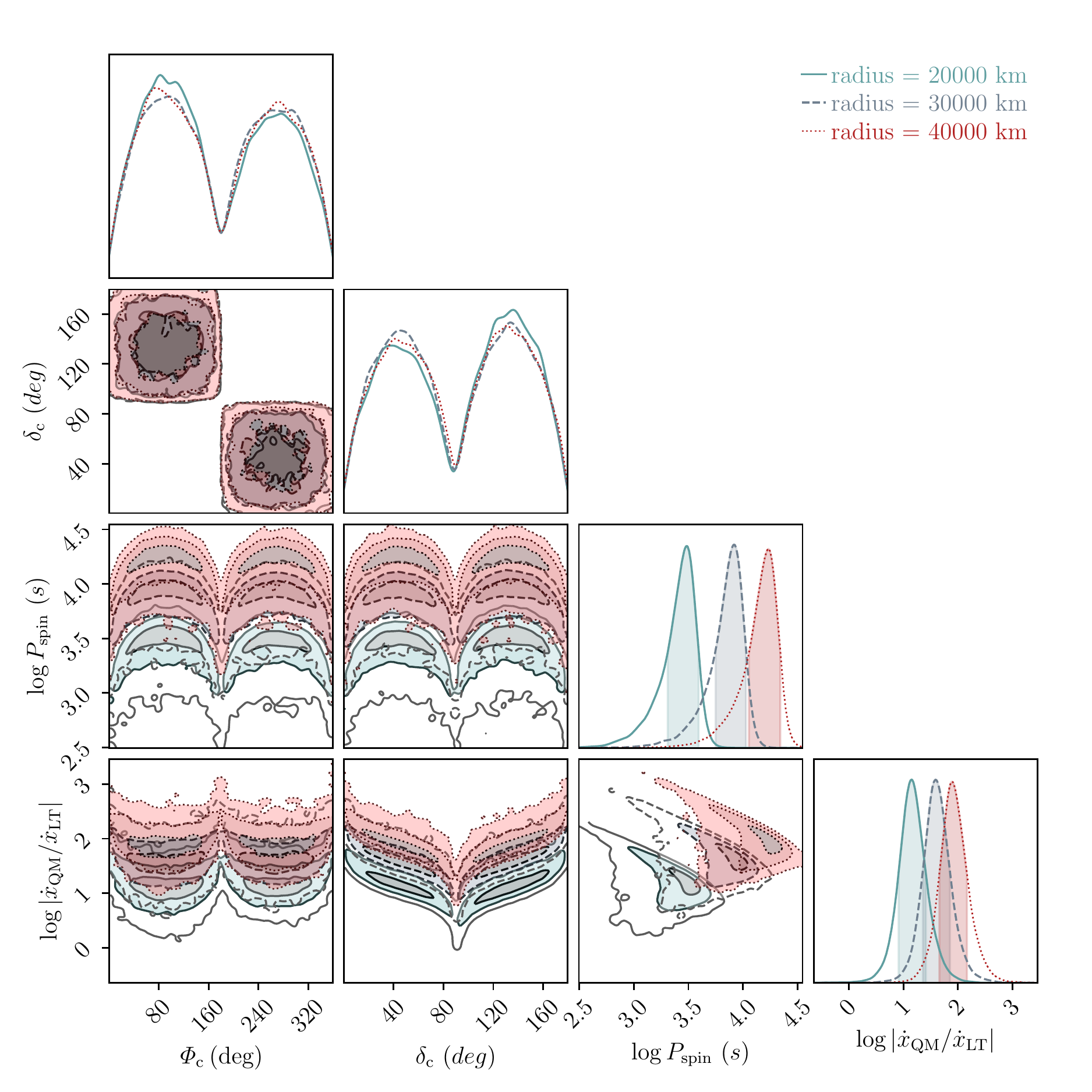}
    \caption{Corner plot showing the constraints on the rotation period of the WD assuming that the measured $\dot{x}$ is due to spin-orbit interactions from the WD. $\phi_{\rm C}$ is the precession phase of the WD spin, and $\delta_{\rm C}$ denotes its misalignment with respect to the orbital angular momentum. The solid blue, dashed olive, and dot-dashed red lines indicate Markov chain Monte Carlo runs assuming a WD radius of 20000, 30000, and 40000 km, respectively. We correspondingly infer a mean rotation rate of 3090, 8472, and 17218 seconds for the WD rotation. The bottom rightmost panel indicates the ratio between the contributions to $\dot{x}$ from classical quadrupole moment (QM) and the relativistic Lense-Thirring precession (LT) of the orbit. For the radius of 20,000 km, we find the LT contribution to be a few percent at most, and for higher radii it is completely negligible. Note that for clarity we have assumed the sense of the inclination here to be $< 90 \deg$. For inclination $> 90 \deg,$ which is also equally likely,  the constraints on the period and the component contributions remain the same, but the constraints on $(\delta_{\rm C}, \Phi_{\rm C})$ change as $\delta_{\rm C} = \delta_{\rm C} -90^{\circ}$ and $\Phi_{\rm C} = \Phi_{\rm C} + 180^{\circ}$.
    }
\label{fig:wdspin}
\end{figure*}

\subsection*{Spin-orbit coupling}

The final effect that could give rise to $\dot{x}_{\rm obs}$ is spin orbit interaction. As mentioned earlier, in a spin-misaligned system, the spin angular momentum of the pulsar and the companion, and the orbital angular momentum precess around the total angular momentum vector. The precession of the orbital plane is induced by two effects: A classical Newtonian quadrupole moment due to the oblateness of the star (QPM) and a relativistic frame-dragging effect termed LT precession (\citealt{LenseThirring1918,Barker&Oconnel1975,DamourTaylor1992}).

In the following equations we use $\mathbb{A}$ to denote the rotating star under consideration (either the pulsar or the WD) and $\mathbb{B}$, its companion. The variables that do not have subscripts denote the pulsar's orbital parameters unless explicitly specified otherwise. The contribution of QPM ($\dot{x}_{\rm QPM}$) is given by
\begin{equation}
\label{eq:QPM}
  \dot{x}_{\rm QPM} = x\left(\frac{2\pi}{P_{\rm B}}\right) Q \cot\textit{i} 
    \sin2\delta_{\mathbb{A}} \sin\Phi^0_{\mathbb{A}},
\end{equation}
where 
\begin{equation}
\label{eq:QPM2}
  Q = \frac{k_2 R^2_{\mathbb{A}} \hat{\Omega}_{\mathbb{A}}^2}{2a^2(1- e^2)^2} \quad {\rm with} \quad  
 \hat{\Omega}_{\mathbb{A}} \equiv \frac{\Omega_{\mathbb{A}}}{(GM_{\mathbb{A}}/R_{\mathbb{A}}^3)^{1/2}}
\end{equation} 
\noindent and $\Omega_{\mathbb{A}} = 2\pi/P_{\mathbb{A}}$, where $a$, $P_\mathbb{A}, \delta_\mathbb{A},\Phi_\mathbb{A}^{0},M_{\mathbb{A}},R_{\mathbb{A}}$ and $k_2$ denote the orbital separation, spin period, spin-misalignment angle, precession phase at time $t=T_0$, mass, radius, and the apsidal motion constant of star $\mathbb{A}$, respectively \citep{SmarrBlandford1976,Lai95,Wex98}. A definition of the angles and vectors used throughout this appendix is provided in Fig. \ref{fig:geometry}. Here we have assumed the precession phase $\Phi_\mathbb{A}\sim \Phi_\mathbb{A}^{0}$ as $\Omega_{\rm geod}^{\mathbb{A}}$ is very small.

The contribution to $\dot{x}_{\rm obs}$ from LT precession ($\dot{x}_{\rm LT}$) is given by 
\begin{equation}
 \label{eq:LT}
 \dot{x}_{\rm LT} \simeq - x\frac{GS_{\mathbb{A}}}{c^2 a^3 (1-e^2)^{3/2}}\left(2 + \frac{3M_{\mathbb{B}}}{2M_{\mathbb{A}}}\right) \cot \textit{i} \sin{\delta_{\mathbb{A}}} \sin{\Phi^0_{\mathbb{A}}}, 
 \end{equation} 
 \noindent where $S_{\mathbb{A}} = I_{\mathbb{A}}\Omega_{\mathbb{A}}$ is the spin angular momentum, with $I_{\mathbb{A}}$ being the moment of inertia of the body ${\mathbb{A}}$ \citep{DamourTaylor1992}. Using Eqs. \ref{eq:QPM},  \ref{eq:QPM2}, and \ref{eq:LT}, we compute the expected contributions due to the rotation of the pulsar, $\dot{x}_{\rm SO}^{\rm P}$, and of the companion, $\dot{x}_{\rm SO}^{\rm C}$.

Assuming a nominal moment of inertia of $1.27 \times 10^{38}~\rm  kg~m^2$ and a radius of 10\,km for the pulsar, we find the contribution from QPM ($\dot{x}_{\rm QPM}^{\rm P}$) to be of the order of $10^{-34}$ and hence irrelevant. We also obtain a maximum of $\sim 2 \times 10^{-16}$ for $\dot{x}_{\rm LT}^{\rm P},$ which is at most 5\% of the maximum likelihood value of the measurement. 
 
In order to estimate the effect of the spin of the WD companion, we  assumed a nominal $k_{2}=0.1$ and that the moment of inertia can be computed as $I=0.2 M_{\rm C} {R_{\rm C}^2}$ and performed a Markov chain Monte Carlo computation to explore the parameter space of $\delta_{\rm C},$ and $\theta_{\rm C},$ and thus constrain the spin period of the WD that could give rise to the observed $\dot{x}_{\rm obs}$. The approach is similar to that presented in \cite{VenkatramanKrishnanEtAl2020b}. Given the uncertainty in the WD radius (see Sect. \ref{subsec:wdmodels}), we performed the computations assuming radii of 20000, 30000, and 40000 km, respectively. Fig.\,\ref{fig:wdspin} shows the posterior distributions of $\delta_{\rm C},$ and $\theta_{\rm C}$ and the absolute ratio of the contributions originating from QPM and LT. We find that $\dot{x}_{\rm obs}$ can be almost entirely ascribed to QPM for rotational periods of the WD of the order of a few hours, not uncommon for a WD in a millisecond pulsar binary. 

In summary, the bulk of $\dot{x}_{\rm obs}$ can be caused by the quadrupole moment caused by the spin of the WD companion. The spin of the pulsar is also likely misaligned, however not so much that all of $\dot{x}_{\rm obs}$ is attributable to the changing aberration. The misalignment of the pulsar gives rise to a combined contribution from LT precession and changing aberration at the few percent level.  However, this still leaves the puzzling origin of the misalignment between the orbital angular momentum and the spin axis of at least one of the two bodies in the binary, which is needed by both most viable models discussed above.


\end{document}